\documentclass[fleqn,twoside]{article}
\usepackage{espcrc1}

\newcommand{\AmS}{{\protect\the\textfont2
  A\kern-.1667em\lower.5ex\hbox{M}\kern-.125emS}}

\title{Towards balanced clustering - part 1 (preliminaries)}

\author{Mark Sh. Levin
\thanks{
 Mark Sh. Levin:~
 Inst. for Inform. Transmission Problems,
 Russian Academy of Sciences;
  http://www.mslevin.iitp.ru;
 email: mslevin@acm.org
  } }

\begin{document}

\maketitle

\begin{abstract}
 The article contains a preliminary glance
 at balanced clustering  problems.
 Basic balanced structures and combinatorial balanced problems
 are briefly described.
 A special attention is targeted to various balance/unbalance
 indices (including some new versions of the indices):
 by cluster cardinality,
 by cluster weights,
 by inter-cluster edge/arc weights,
 by cluster element structure
 (for element multi-type clustering).
 Further,
 versions of optimization clustering problems are suggested
 (including multicriteria problem formulations).
 Illustrative numerical examples describe
 calculation of balance indices and
 element multi-type balance clustering problems
 (including example for design of student teams).

~~


\vspace{1pc}
\end{abstract}

\maketitle

\tableofcontents

\newcounter{cms}
\setlength{\unitlength}{1mm}

%
\section{Introduction}

 Balancing processes play central roles in many theoretical
 and practical fields (Fig. 1)
 \cite{andres08,arg04,bader13,bich13,bori14,boy07,cart56,dai16,davis67,dolg06,khan14,knu97,liux12,mus06b,rais08,rob78,shang92,yild13}.
 The corresponding balancing problems are basic ones
 in various engineering domains, for example:
 manufacturing systems,
 computing systems,
 power/electricity systems,
 radio engineering systems,
 communication systems, and
 civil engineering systems.
 Similar balancing problems are examined in
 engineering management (e.g., coordination science),
 communications (e.g., synchronization as time-based balancing),
 organization science, psychology.
 In recent decades,
 the balancing approaches are very significant from the viewpoint
 of system modularity
 \cite{bald00,eth04,jose05,lev15,ma16b}.

\begin{center}
\begin{picture}(105,32)
\put(20,00){\makebox(0,0)[bl]{Fig. 1.
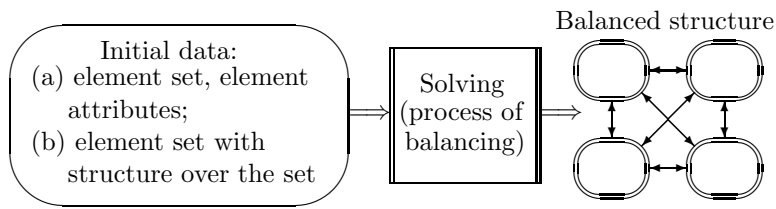
 Scheme of balancing process}}
\put(22.5,17){\oval(45,24)}

\put(12,24.5){\makebox(0,0)[bl]{Initial data:}}
\put(2.8,20.2){\makebox(0,0)[bl]{(a) element set, element}}
\put(7.7,16){\makebox(0,0)[bl]{attributes;}}
\put(2.8,11.5){\makebox(0,0)[bl]{(b) element set with }}
\put(7.7,08.1){\makebox(0,0)[bl]{structure over the set}}

\put(45,16.5){\makebox(0,0)[bl]{\(\Longrightarrow\)}}

\put(50.5,08){\line(1,0){20}} \put(50.5,26){\line(1,0){20}}
\put(50.5,08){\line(0,1){18}} \put(70.5,08){\line(0,1){18}}
\put(51,08){\line(0,1){18}} \put(70,08){\line(0,1){18}}

\put(54.5,19.5){\makebox(0,0)[bl]{Solving}}
\put(51.7,15.5){\makebox(0,0)[bl]{(process of}}
\put(52.5,11.5){\makebox(0,0)[bl]{balancing)}}

\put(70.5,16.5){\makebox(0,0)[bl]{\(\Longrightarrow\)}}

\put(72.5,28.5){\makebox(0,0)[bl]{Balanced structure}}

\put(80,23){\oval(10,08)} \put(95,23){\oval(10,08)}
\put(80,23){\oval(09,07)} \put(95,23){\oval(09,07)}

\put(85,23){\vector(1,0){05}} \put(90,23){\vector(-1,0){05}}

\put(80,10){\oval(10,08)} \put(95,10){\oval(10,08)}
\put(80,10){\oval(09,07)} \put(95,10){\oval(09,07)}

\put(85,10){\vector(1,0){05}} \put(90,10){\vector(-1,0){05}}

\put(80,14){\vector(0,1){05}} \put(80,19){\vector(0,-1){05}}
\put(95,14){\vector(0,1){05}} \put(95,19){\vector(0,-1){05}}

\put(84,13){\vector(1,1){07}} \put(91,20){\vector(-1,-1){07}}
\put(84,20){\vector(1,-1){07}} \put(91,13){\vector(-1,1){07}}

\end{picture}
\end{center}

 The basic balancing problem consists in
 partitioning the element set
 (while taking into account element parameters,
 structure over the elements)
 into interconnected element groups (clusters)
 which are balanced (by cardinality of cluster elements,
 weight of cluster as a total weight of cluster elements,
 weight of cluster interconnections, structure of cluster, etc.).
 A general framework of balancing problems domain is depicted in Fig. 2.

\begin{center}
\begin{picture}(110,52)
\put(09.3,00){\makebox(0,0)[bl]{Fig. 2.
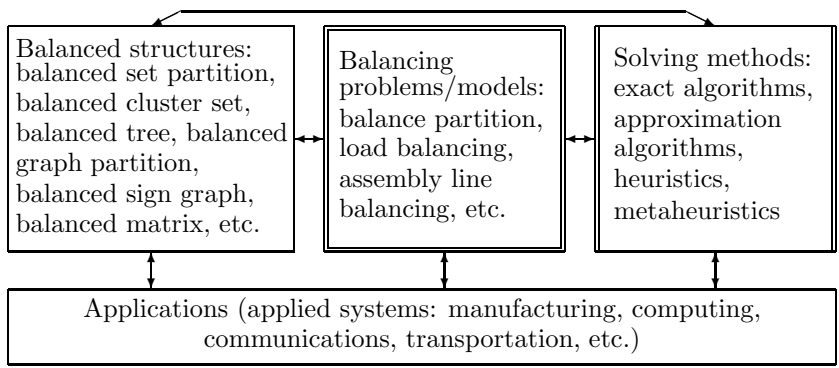
 General framework of balancing problems domains}}

\put(00,05){\line(1,0){110}} \put(00,15){\line(1,0){110}}
\put(00,05){\line(0,1){10}} \put(110,05){\line(0,1){10}}

\put(10,10.5){\makebox(0,0)[bl]{Applications
 (applied systems:  manufacturing, computing,}}

\put(26,06.5){\makebox(0,0)[bl]{communications,
 transportation, etc.)}}

\put(19,15){\vector(0,1){5}} \put(19,20){\vector(0,-1){5}}
\put(58,15){\vector(0,1){5}} \put(58,20){\vector(0,-1){5}}
\put(94,15){\vector(0,1){5}} \put(94,20){\vector(0,-1){5}}

\put(00,20){\line(1,0){38}} \put(00,50){\line(1,0){38}}
\put(00,20){\line(0,1){30}} \put(38,20){\line(0,1){30}}

\put(01,46){\makebox(0,0)[bl]{Balanced structures:}}
\put(01,42){\makebox(0,0)[bl]{balanced set partition,}}
\put(01,38){\makebox(0,0)[bl]{balanced cluster set, }}
\put(01,34){\makebox(0,0)[bl]{balanced tree, balanced}}
\put(01,30){\makebox(0,0)[bl]{graph partition,}}
\put(01,26){\makebox(0,0)[bl]{balanced sign graph,}}
\put(01,22){\makebox(0,0)[bl]{balanced matrix, etc.}}

\put(38,35){\vector(1,0){4}} \put(42,35){\vector(-1,0){4}}

\put(42,20){\line(1,0){32}} \put(42,50){\line(1,0){32}}
\put(42,20){\line(0,1){30}} \put(74,20){\line(0,1){30}}

\put(42.5,20.5){\line(1,0){31}} \put(42.5,49.5){\line(1,0){31}}
\put(42.5,20.5){\line(0,1){29}} \put(73.5,20.5){\line(0,1){29}}

\put(44,43.6){\makebox(0,0)[bl]{Balancing}}
\put(44,40){\makebox(0,0)[bl]{problems/models:}}
\put(44,36){\makebox(0,0)[bl]{balance partition,}}
\put(44,32){\makebox(0,0)[bl]{load balancing,}}
\put(44,28){\makebox(0,0)[bl]{assembly line}}
\put(44,24){\makebox(0,0)[bl]{balancing, etc.}}

\put(23,52){\vector(-2,-1){4}} \put(90,52){\vector(2,-1){4}}
\put(23,52){\line(1,0){67}}


\put(78,20){\line(1,0){32}} \put(78,50){\line(1,0){32}}
\put(78,20){\line(0,1){30}} \put(110,20){\line(0,1){30}}
\put(78.5,20){\line(0,1){30}} \put(109.5,20){\line(0,1){30}}

\put(80.5,44){\makebox(0,0)[bl]{Solving methods:}}
\put(80.5,40){\makebox(0,0)[bl]{exact algorithms,}}
\put(80.5,36){\makebox(0,0)[bl]{approximation}}
\put(80.5,32){\makebox(0,0)[bl]{algorithms,}}
\put(80.5,28){\makebox(0,0)[bl]{heuristics,}}
\put(80.5,24.4){\makebox(0,0)[bl]{metaheuristics}}

\put(78,35){\vector(-1,0){4}} \put(74,35){\vector(1,0){4}}

\end{picture}
\end{center}

 It is reasonable to point out the following
  main application domains with using the balanced structures:

 {\it 1.} hierarchical organization of storage and search processes in
 information systems and in computer systems
 (e.g., various balanced search trees BSTs:
 AVL trees, red-black trees, splay trees)
  \cite{adel62,baer77,karl76,knu97,lev15,lev16dafi,sle85};

 {\it 2.} balanced parallel scheduling of computing processes
 (e.g., computing in multi-processor systems, distributed computing)
  \cite{bader13,bich13,tsail92};

 {\it 3.} service partitioning in a grid environment
  \cite{mus06a,mus06b};

 {\it 4.} balanced partitioning in networks
  (e.g., modularization),
  balanced allocation of centers in networks
  \cite{bari93,dai16};

 {\it 5.} balanced partitioning of algorithms \cite{shang92};

 {\it 6.} balanced manufacturing scheduling in manufacturing
 systems
 (e.g., assembly line balancing problems)
    \cite{amen06,andres08,bori14,boy07,dolg06,dolg06a,erel98,hoff90};

 {\it 7.} design of balanced hierarchical structures in organizations
   \cite{arg04,rais08};

 {\it 8.} design and management in communication systems
  (network design, routing, etc.)
  based on balanced hierarchical structures
  \cite{khan14,liux12};
  and

 {\it 9.} design of distributed defence systems
    \cite{fuy14,yild13}.

 Evidently, balanced clustering problems
 (including combinatorial clustering, constrained clustering,
 etc.)
 are used as the basic balancing combinatorial models for all domains
 \cite{basu08,borg16,lev15c,lev15d}.
 Some basic balanced clustering problems are pointed out in Table 1.

 In general, the clustering problem is the following.
 Let \(A = \{ a_{1},...,a_{j},...,a_{n}\}\) be the initial set of
 elements (items, objects).
 Usually, the following characteristics are examined:

 (1) parameters of each element \(a_{j} \in A\) as vector
 \(\overline{p}(a_{j})=(p_{1}(a_{j}),...,p_{i}(a_{j}),...,p_{m}(a_{j}))\);

 (2) structure(s) (binary relation(s)) over the elements set \(A\):
   \(G=(A,E)\)
   where \(E\) is a set of edges/arcs (or weighted edges/arcs).

   A clustering solution consists of a set of clusters
  (e.g., without intersections)
   \cite{lev15c,lev15d}:
 \[ \widetilde{X} = \{X_{1},..., X_{\iota},...,X_{\lambda} \}, \]
 i.e., dividing the set \(A\) into clusters:
 \(X_{\iota} \subseteq A\)
 \(\forall \iota = \overline{1,\lambda}\),
 \(\eta_{\iota} = |X_{\iota}|\) is the cluster size
 (cardinality of cluster \(X_{\iota}\),
 \(\iota \in  \overline{1,\lambda} \) ).

\begin{center}
 {\bf Table 1.} Basic balanced clustering problems\\
\begin{tabular}{| c | l| l|l |}
\hline
 No.&Basic types of initial data& Results&Some source(s)\\

\hline
 1.&Set of elements, element&Set of
    balanced clusters
       &\cite{basu08,borg16,kui13,lev15c,liao13}\\

  &parameters& &\cite{low08,shang10,shap67,zhao15}\\

 2.&Set of elements with relations:&& \cite{abbasi07,bala94,cheng11,current86,gupta03,kui14a}  \\
    &&& \cite{lev12hier,lev15,obr10,pirkul91}\\
    &&& \cite{sim05,youn04,youn06,zhao15} \\

 2.1.&Set of elements with precedence&(1) chain over balanced clusters&\\
   &relation  &(balanced one-processor scheduling)&\\

   & &(2) parallel balanced element groups &\\
   & &(precedence over elements in each &\\
   & & group, multiprocessor scheduling)&\\

 2.2.&Set of elements with various&(1) Tree/forest/hierachy over&\\

   &binary relation(s)  & balanced clusters (search structure,
   &\\

   &(e.g., precedence, inclusion)& hierarchical storage paging, &\\
    && balanced broadcasting)&\\

   &&(2) \(k\)-layer network architecture
    &\\

     &&(hierarchical communications,&\\
   &&hierarchical distributed computing)&\\

 3.&Set of element chains&Special time interval balanced
                         &\cite{bori14}\\
   &(trajectories)    &scheduling of element chains&\\
   &&(scheduling in  homebuilding)&\\

 4.&Set of element structures (e.g.,&Set of balanced clusters of element
                                      &\cite{lev15c,lev16dafi}\\
   &programs, data structures, groups&structures (e.g., distributed &\\
   &of technological operations, teams) &computing, technological plans)&\\

\hline
\end{tabular}
\end{center}

 The basic balance (or unbalance/mismatch)
 index (parameter) for clustering solution is
 the difference between the maximal cardinality of a cluster and
 minimal cardinality of a cluster
 (in the considered clustering solution):
 \[ B^{c}( \widetilde{X} ) =
  \max_{ \iota = \overline{1,\lambda} } \eta_{\iota} -
 \min_{ \iota = \overline{1,\lambda} } \eta_{\iota} .\]
 Thus, the balanced clustering problem is targeted to search for
 the clustering solution with minimum balance index
 (e.g., index above).
 Clearly,
 additions to the balanced problem statement can involve the
 following:
 (a) some constraints:
 the fixed number of clusters,
 restriction(s) to cluster sizes (including specified integer
 interval), etc.;
 (b) objective function(s)
 (e.g., minimization of total interconnections weight between clusters,
 maximization of total element connections weight in clusters).

 This material contains an author preliminary outline  of balanced
 clustering problems.
 The basic balanced structures are pointed out.
 New balance/unbalance indices for clustering solutions are suggested.
 Formulations of balanced optimization problems are described
 (including multicriteria problem formulations).
 Several numerical examples illustrate
 calculation of balance/unbalance indices for clustering
 solutions.
 A special example for design of student teams
 (as element multi-type cluster structure balanced clustering problem)
 is described.
 The material can be considered as a continuation
 of the author preprint on combinatorial
 clustering  \cite{lev15c} and
 the corresponding article \cite{lev15d}.

\newpage
\section{Basic balanced structures}

 The following basic balanced structures can be pointed out:

 {\bf 1.} Balanced set partition
 (by cardinality, by element weights, element structure;
 in the general case, the obtained subsets can have intersections):
  (a) a basic  illustration (Fig. 3),
  (b) balancing by element structure
     in each subset (Fig. 4).

 {\bf 2}. Chain of balanced clusters
 (balanced manufacturing line,
 balanced chain of computing tasks groups, etc.)
 (Fig. 5).

 {\bf 3.}  Balanced packing of bins (e.g., multi-processor scheduling) (Fig. 6).

 {\bf 4}. Balanced \(\beta\)-layer clustering
  (i.e., partitioning the initial element set into clusters and layers;
  e.g., to obtain a two-layer hierarchy
  with balances at the layer of clusters and at the layer of
  cluster groups, this is a two-balancing case)
   (Fig. 7).

 {\bf 5.} Balanced trees \cite{aho83,cormen90,knu97}:~
 {\it 5.1.} balanced trees: by height (Fig. 8), by degree
 (Fig. 9),

 {\it 5.2.} balanced search trees (as search index structures):
   (a) AVL trees, self-balancing (or height balancing) binary search tree
   \cite{adel62},
   (b) \(B\)-trees (path length from the root to each leaf equals \(b\) or   \(b+1\)
   (where \(b\) is a constant)
     (Fig. 8) \cite{bayer72a,cormen90,knu97},
  (c) Red-Black trees (symmetric binary B-tree)
  \cite{bayer72,cormen90}, etc.

 {\bf 6.} Balanced graph partition
 (e.g., partition of a graph into balanced subgraphs/communities
 while taking into account vertex weights or/and edge weights) (Fig. 10).

 {\bf 7.} \(k\)-optimal partition of directed graph (path partition)
         \cite{bena06,berg82,berger08} (Fig. 11).

 {\bf 8.} Balanced
  signed graph \cite{aki81,gar79,har53,rob76,rob78}
 (Fig. 12).

 {\bf 9.} Multi-layer structure of balanced clusters in networking:
 {\it 9.1} hierarchy over balanced clusters (networking) (Fig. 13);
 {\it 9.2}  balanced clustering based multi-layer network structure (Fig. 14).

 {\bf 10.} Balanced matrices \cite{berg72,conf05,rayan88}.

~~

\begin{center}
\begin{picture}(67,42)
\put(03.5,00){\makebox(0,0)[bl]{Fig. 3. Balanced set partition}}

\put(06.5,18){\oval(13,26)}

\put(01.5,20){\makebox(0,0)[bl]{Initial}}
\put(0.5,16){\makebox(0,0)[bl]{element}}
\put(04.5,12){\makebox(0,0)[bl]{set}}

\put(14,24){\vector(3,1){5}} \put(14,18){\vector(1,0){5}}
\put(14,12){\vector(3,-1){5}}

\put(15,38){\makebox(0,0)[bl]{Balanced subsets (by}}
\put(12,35){\makebox(0,0)[bl]{(cardinalities, by element}}
\put(08,32){\makebox(0,0)[bl]{weights, by element structures)}}

\put(25,11){\oval(10,12)} \put(24,24){\oval(08,12)}
\put(37,27){\oval(16,08)} \put(42,21){\oval(12,08)}
\put(39,11){\oval(14,10)}

\end{picture}
%
\begin{picture}(79,30)

\put(00,00){\makebox(0,0)[bl]{Fig. 4.
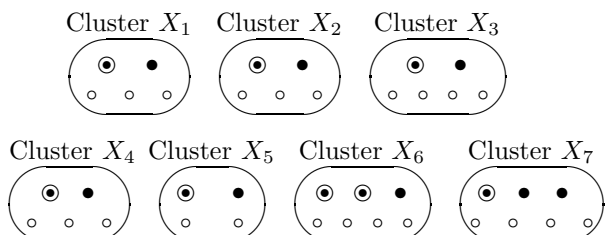
 Balanced cluster set (by subset structures)}}

\put(00,15){\makebox(0,0)[bl]{Cluster \(X_{4}\)}}

\put(08,09.5){\oval(16,10)}

\put(05.5,11){\circle*{1.1}} \put(05.5,11){\circle{2.0}}

\put(10.5,11){\circle*{1.3}}

\put(03,07){\circle{1.2}} \put(08,07){\circle{1.2}}
\put(13,07){\circle{1.2}}

\put(18.5,15){\makebox(0,0)[bl]{Cluster \(X_{5}\)}}

\put(27,09.5){\oval(14,10)}

\put(23.5,11){\circle*{1.1}} \put(23.5,11){\circle{2.0}}

\put(30.5,11){\circle*{1.3}}

\put(23.5,07){\circle{1.2}} \put(30.5,07){\circle{1.2}}

\put(38.5,15){\makebox(0,0)[bl]{Cluster \(X_{6}\)}}

\put(47,09.5){\oval(18,10)}

\put(42,11){\circle*{1.1}} \put(42,11){\circle{2.0}}
\put(47,11){\circle*{1.1}} \put(47,11){\circle{2.0}}

\put(52,11){\circle*{1.3}}

\put(41,07){\circle{1.2}} \put(45,07){\circle{1.2}}
\put(49,07){\circle{1.2}} \put(53,07){\circle{1.2}}

\put(61,15){\makebox(0,0)[bl]{Cluster \(X_{7}\)}}

\put(69.5,09.5){\oval(19,10)}

\put(63.5,11){\circle*{1.1}} \put(63.5,11){\circle{2.0}}

\put(68.5,11){\circle*{1.3}} \put(73.5,11){\circle*{1.3}}

\put(62,07){\circle{1.2}} \put(67,07){\circle{1.2}}
\put(72,07){\circle{1.2}} \put(77,07){\circle{1.2}}

\put(07.5,32){\makebox(0,0)[bl]{Cluster \(X_{1}\)}}

\put(16,26.5){\oval(16,10)}

\put(013,28){\circle*{1.1}} \put(013,28){\circle{2.0}}

\put(19,28){\circle*{1.3}}

\put(11,24){\circle{1.2}} \put(016,24){\circle{1.2}}
\put(21,24){\circle{1.2}}

\put(27.5,32){\makebox(0,0)[bl]{Cluster \(X_{2}\)}}

\put(36,26.5){\oval(16,10)}

\put(33,28){\circle*{1.1}} \put(33,28){\circle{2.0}}

\put(39,28){\circle*{1.3}}

\put(31,24){\circle{1.2}} \put(36,24){\circle{1.2}}
\put(41,24){\circle{1.2}}

\put(48.5,32){\makebox(0,0)[bl]{Cluster \(X_{3}\)}}
\put(57,26.5){\oval(18,10)}

\put(54,28){\circle*{1.1}} \put(54,28){\circle{2.0}}

\put(60,28){\circle*{1.3}}

\put(51,24){\circle{1.2}} \put(55,24){\circle{1.2}}
\put(59,24){\circle{1.2}} \put(63,24){\circle{1.2}}

\end{picture}
\end{center}



\begin{center}
\begin{picture}(77,34)

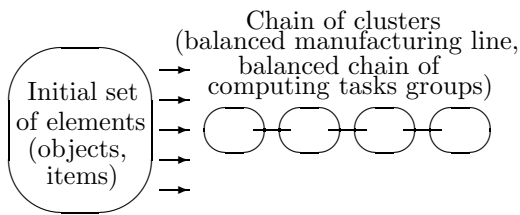
\put(06,00){\makebox(0,0)[bl]{Fig. 5. Chain of balanced
 clusters}}

\put(09.5,16){\oval(19,22)}

\put(2.5,20){\makebox(0,0)[bl]{Initial set}}
\put(1,16){\makebox(0,0)[bl]{of elements}}
\put(02.5,11.5){\makebox(0,0)[bl]{(objects,}}
\put(05,08){\makebox(0,0)[bl]{items)}}

\put(20,24){\vector(1,0){04}} \put(20,20){\vector(1,0){04}}
\put(20,16){\vector(1,0){04}} \put(20,12){\vector(1,0){04}}
\put(20,08){\vector(1,0){04}}

\put(31.5,29.5){\makebox(0,0)[bl]{Chain of clusters}}
\put(22,26){\makebox(0,0)[bl]{(balanced manufacturing line,}}
\put(30.5,23.5){\makebox(0,0)[bl]{balanced chain of}}
\put(26.5,20){\makebox(0,0)[bl]{computing tasks groups)}}

\put(30,16){\oval(08,6)}

\put(40,16){\oval(08,6)}

\put(50,16){\oval(08,6)}

\put(60,16){\oval(08,6)}

\put(32.5,16){\line(1,0){05}} \put(42.5,16){\line(1,0){05}}
\put(52.5,16){\line(1,0){05}}

\end{picture}
%
\begin{picture}(66,42)
\put(06,00){\makebox(0,0)[bl]{Fig. 6.
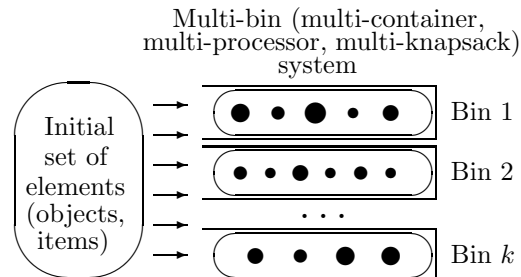
  Balanced packing of bins}}

\put(08.5,18){\oval(17,26)}

\put(04,24){\makebox(0,0)[bl]{Initial}}
\put(04,20){\makebox(0,0)[bl]{set of}}
\put(1.5,16){\makebox(0,0)[bl]{elements}}
\put(1.5,11.5){\makebox(0,0)[bl]{(objects,}}
\put(03,08){\makebox(0,0)[bl]{items)}}

\put(18.5,28){\vector(1,0){04.5}}
\put(18.5,24){\vector(1,0){04.5}}
\put(18.5,20){\vector(1,0){04.5}}
\put(18.5,16){\vector(1,0){04.5}}
\put(18.5,12){\vector(1,0){04.5}}
\put(18.5,08){\vector(1,0){04.5}}

\put(21,37.5){\makebox(0,0)[bl]{Multi-bin (multi-container,}}
\put(17,34.5){\makebox(0,0)[bl]{multi-processor, multi-knapsack)}}
\put(35,31.5){\makebox(0,0)[bl]{system}}

\put(41,27){\oval(29,6)}

\put(58,26){\makebox(0,0)[bl]{Bin \(1\)}}

\put(25,23.5){\line(1,0){31}} \put(25,30.5){\line(1,0){31}}
\put(56,23.5){\line(0,1){07}}

\put(30,27){\circle*{2.4}} \put(35,27){\circle*{1.8}}
\put(40,27){\circle*{2.7}} \put(45,27){\circle*{1.4}}
\put(50,27){\circle*{2.0}}

\put(41,19){\oval(29,6)}

\put(58,18){\makebox(0,0)[bl]{Bin \(2\)}}

\put(25,15.5){\line(1,0){31}} \put(25,22.5){\line(1,0){31}}
\put(56,15.5){\line(0,1){07}}

\put(30,19){\circle*{1.7}} \put(34,19){\circle*{1.5}}
\put(38,19){\circle*{2.1}} \put(42,19){\circle*{1.4}}
\put(46,19){\circle*{1.6}} \put(50,19){\circle*{1.3}}

\put(38,13){\makebox(0,0)[bl]{{\bf .~.~.} }}

\put(41,08){\oval(29,6)}

\put(58,07){\makebox(0,0)[bl]{Bin \(k\)}}

\put(25,4.5){\line(1,0){31}} \put(25,11.5){\line(1,0){31}}
\put(56,4.5){\line(0,1){07}}

\put(32,08){\circle*{2.2}} \put(38,08){\circle*{1.9}}
\put(44,08){\circle*{2.3}} \put(50,08){\circle*{2.5}}

\end{picture}
\end{center}
%



%
\begin{center}
\begin{picture}(74,35)

\put(01.5,00){\makebox(0,0)[bl]{Fig. 7.
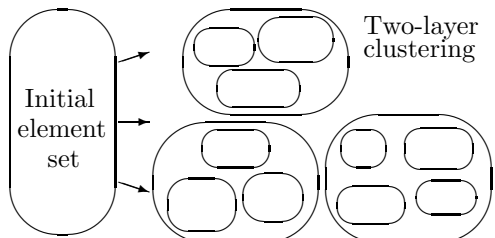
 Balanced two-layer clustering}}

\put(07,20){\oval(14,30)}

\put(02,22){\makebox(0,0)[bl]{Initial}}
\put(01,18){\makebox(0,0)[bl]{element}}
\put(05,14){\makebox(0,0)[bl]{set}}

\put(14.5,28){\vector(3,1){4}} \put(14.5,20){\vector(1,0){4}}
\put(14.5,12){\vector(3,-1){4}}

\put(25.5,09){\oval(09,07)} \put(35,10){\oval(08,06.5)}
\put(30,16.5){\oval(09,05)}

 \put(30,12){\oval(22,16)}

\put(48,09){\oval(09,05)} \put(58,10){\oval(08,04.5)}
\put(47,16.5){\oval(06,05)} \put(57,16.5){\oval(09,05.5)}

\put(53,12.5){\oval(22,17)}

\put(28.5,30){\oval(08,05)} \put(38,31){\oval(10,06)}
\put(33,24.5){\oval(11,05)}

\put(34,28){\oval(22,14)}

\put(47,31){\makebox(0,0)[bl]{Two-layer}}
\put(47,28){\makebox(0,0)[bl]{clustering}}

\end{picture}
%
\begin{picture}(66,26)

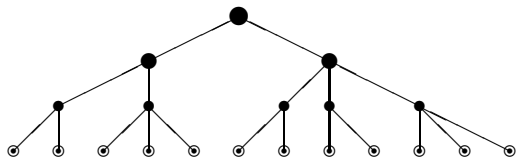
\put(010,00){\makebox(0,0)[bl]{Fig. 8. Height-balanced tree}}

\put(30,24){\circle*{2.4}}

\put(18,18){\line(2,1){12}} \put(42,18){\line(-2,1){12}}

\put(18,18){\circle*{2.0}}

\put(42,18){\circle*{2.0}}

\put(06,12){\line(2,1){12}} \put(18,12){\line(0,1){06}}

\put(36,12){\line(1,1){6}} \put(42,12){\line(0,1){06}}
\put(54,12){\line(-2,1){12}}

\put(06,12){\circle*{1.4}}

\put(18,12){\circle*{1.4}}

\put(36,12){\circle*{1.4}}

\put(42,12){\circle*{1.4}}

\put(54,12){\circle*{1.4}}

\put(00,06){\line(1,1){06}} \put(06,06){\line(0,1){06}}

\put(12,06){\line(1,1){06}} \put(18,06){\line(0,1){06}}
\put(24,06){\line(-1,1){06}}

\put(30,06){\line(1,1){06}} \put(36,06){\line(0,1){06}}

\put(42,06){\line(0,1){06}} \put(48,06){\line(-1,1){06}}

\put(54,06){\line(0,1){06}} \put(60,06){\line(-1,1){06}}
\put(66,06){\line(-2,1){12}}

\put(00,06){\circle*{0.8}} \put(00,06){\circle{1.4}}

\put(06,06){\circle*{0.8}} \put(06,06){\circle{1.4}}

\put(12,06){\circle*{0.8}} \put(12,06){\circle{1.4}}

\put(18,06){\circle*{0.8}} \put(18,06){\circle{1.4}}

\put(24,06){\circle*{0.8}} \put(24,06){\circle{1.4}}

\put(30,06){\circle*{0.8}} \put(30,06){\circle{1.4}}

\put(36,06){\circle*{0.8}} \put(36,06){\circle{1.4}}

\put(42,06){\circle*{0.8}} \put(42,06){\circle{1.4}}

\put(48,06){\circle*{0.8}} \put(48,06){\circle{1.4}}

\put(54,06){\circle*{0.8}} \put(54,06){\circle{1.4}}

\put(60,06){\circle*{0.8}} \put(60,06){\circle{1.4}}

\put(66,06){\circle*{0.8}} \put(66,06){\circle{1.4}}

\end{picture}
\end{center}



\begin{center}
\begin{picture}(60,44)
\put(02,00){\makebox(0,0)[bl]{Fig. 9. \(3\)-degree tree}}

\put(24,24){\circle*{2.9}}

\put(12,18){\line(2,1){12}} \put(24,18){\line(0,1){06}}
\put(30,18){\line(-1,1){6}}

\put(12,18){\circle*{2.2}}

\put(24,18){\circle*{0.8}} \put(24,18){\circle{1.4}}

\put(30,18){\circle*{2.2}}

\put(06,12){\line(1,1){06}} \put(12,12){\line(0,1){06}}
\put(18,12){\line(-1,1){06}}

\put(24,12){\line(1,1){06}} \put(30,12){\line(0,1){06}}
\put(36,12){\line(-1,1){06}}

\put(06,12){\circle*{1.4}}

\put(12,12){\circle*{0.8}} \put(12,12){\circle{1.4}}
\put(18,12){\circle*{0.8}} \put(18,12){\circle{1.4}}


\put(30,12){\circle*{1.4}}

\put(36,12){\circle*{0.8}} \put(36,12){\circle{1.4}}
\put(24,12){\circle*{0.8}} \put(24,12){\circle{1.4}}

\put(00,06){\line(1,1){06}} \put(06,06){\line(0,1){06}}
\put(12,06){\line(-1,1){06}}

\put(24,06){\line(1,1){06}} \put(30,06){\line(0,1){06}}
\put(36,06){\line(-1,1){06}}

\put(00,06){\circle*{0.8}} \put(00,06){\circle{1.4}}

\put(06,06){\circle*{0.8}} \put(06,06){\circle{1.4}}

\put(12,06){\circle*{0.8}} \put(12,06){\circle{1.4}}

\put(24,06){\circle*{0.8}} \put(24,06){\circle{1.4}}

\put(30,06){\circle*{0.8}} \put(30,06){\circle{1.4}}

\put(36,06){\circle*{0.8}} \put(36,06){\circle{1.4}}

\end{picture}
%
\begin{picture}(65,48.5)

\put(04,00){\makebox(0,0)[bl]{Fig. 10. Graph
 partitioning}}

\put(05,20){\line(1,1){10}}

\put(05,20){\line(1,-1){10}}


\put(30,40){\line(1,-2){05}}


\put(25,30){\line(1,0){10}}


\put(15,30){\line(3,-2){15}}

\put(30,20){\line(3,-2){15}}



\put(15,10){\line(1,0){30}}

\put(15,20){\line(1,1){10}}

\put(50,20){\line(-1,2){05}}

\put(50,20){\line(-1,-2){05}}


\put(05,20){\circle*{0.9}} \put(05,20){\circle{1.7}}
\put(02.2,19){\makebox(0,8)[bl]{\(8\)}}
\put(05,20){\line(1,0){10}}

\put(05,10){\circle*{1}} \put(02,11){\makebox(0,8)[bl]{\(13\)}}
\put(05,10){\line(1,1){10}} \put(05,10){\line(1,0){10}}

\put(15,20){\circle*{0.6}} \put(15,20){\circle{1.4}}
\put(15,20){\circle{2.4}}
\put(12.6,20.8){\makebox(0,8)[bl]{\(9\)}}

\put(15,30){\circle*{1}} \put(14.5,31.5){\makebox(0,8)[bl]{\(4\)}}

\put(05,30){\circle*{1}} \put(02.5,29){\makebox(0,8)[bl]{\(3\)}}
\put(05,30){\line(1,0){10}}

\put(10,40){\circle*{0.6}} \put(10,40){\circle{1.4}}
\put(10,40){\circle{2.4}} \put(07,39){\makebox(0,8)[bl]{\(1\)}}
\put(10,40){\line(-1,-2){05}} \put(10,40){\line(1,-2){05}}

\put(15,20){\line(0,1){10}} \put(15,20){\line(0,-1){10}}

\put(15,10){\circle*{0.9}} \put(15,10){\circle{1.7}}
\put(11.2,07){\makebox(0,8)[bl]{\(14\)}}


\put(30,20){\circle*{1}}
\put(31.5,20.6){\makebox(0,8)[bl]{\(10\)}}
\put(15,20){\line(1,0){15}}

\put(25,30){\circle*{0.9}} \put(25,30){\circle{1.7}}
\put(26.8,30.5){\makebox(0,8)[bl]{\(5\)}}

\put(35,30){\circle*{0.9}} \put(35,30){\circle{1.7}}
\put(35,31){\makebox(0,8)[bl]{\(6\)}}

\put(30,20){\line(-1,2){05}} \put(30,20){\line(1,2){05}}

\put(30,40){\circle*{0.6}} \put(30,40){\circle{1.4}}
\put(30,40){\circle{2.4}} \put(31.5,39){\makebox(0,8)[bl]{\(2\)}}
\put(30,40){\line(-1,-2){05}}


\put(30,20){\line(1,0){10}}

\put(40,20){\circle*{0.6}} \put(40,20){\circle{1.4}}
\put(40,20){\circle{2.4}} \put(38.7,24){\makebox(0,8)[bl]{\(11\)}}

\put(50,20){\circle*{0.6}} \put(50,20){\circle{1.4}}
\put(50,20){\circle{2.4}} \put(44.8,21){\makebox(0,8)[bl]{\(12\)}}

\put(40,20){\line(1,0){10}} \put(40,20){\line(1,2){05}}
\put(40,20){\line(1,-2){05}}

\put(45,30){\circle*{1}} \put(46,29){\makebox(0,8)[bl]{\(7\)}}

\put(45,10){\circle*{1}} \put(45,10){\circle{1.7}}
\put(45.7,08){\makebox(0,8)[bl]{\(15\)}}

\put(16.4,5){\makebox(0,0)[bl]{Cluster \(X_{1}'\)}}
\put(09.5,15){\oval(18,18)}

\put(0.7,44){\makebox(0,0)[bl]{Cluster \(X_{2}'\)}}
\put(09,34.5){\oval(18,17)}

\put(22.5,44){\makebox(0,0)[bl]{Cluster \(X_{3}'\)}}
\put(30.5,31){\oval(14,25)}

\put(38,33.6){\makebox(0,0)[bl]{Cluster \(X_{4}'\)}}
\put(45.5,20){\oval(14,26)}

\end{picture}
\end{center}

~~

\begin{center}
\begin{picture}(82,33.5)\

\put(04,00){\makebox(0,0)[bl]{Fig. 11. \(k\)-optimal partition
 of digraph}}

\put(25.5,29.5){\makebox(0,0)[bl]{\(G=(A,\overrightarrow{E})\)}}
\put(33.5,17){\oval(47,24)}

\put(45,24.5){\makebox(0,0)[bl]{\(\mu_{p}\)}}

\put(44,23){\circle*{1.3}} \put(49,23){\circle*{1.3}}
\put(49,18){\circle*{1.3}} \put(49,13){\circle*{1.3}}
\put(49,08){\circle*{1.3}}

\put(49,08){\vector(0,1){04}} \put(49,13){\vector(0,1){04}}
\put(49,18){\vector(0,1){04}} \put(49,23){\vector(-1,0){04}}

\put(33,24.5){\makebox(0,0)[bl]{\(\mu_{p-1}\)}}

\put(37,23){\circle{1.6}} \put(42,18){\circle{1.6}}
\put(42,13){\circle{1.6}} \put(42,08){\circle{1.6}}

\put(37,23){\circle*{0.8}} \put(42,18){\circle*{0.8}}
\put(42,13){\circle*{0.8}} \put(42,08){\circle*{0.8}}

\put(42,08){\vector(0,1){04}} \put(42,13){\vector(0,1){04}}
\put(42,18){\vector(-1,1){04.5}}

\put(30,18){\makebox(0,0)[bl]{{\bf .~.~.}}}
\put(30,12){\makebox(0,0)[bl]{{\bf .~.~.}}}

\put(16,19.5){\makebox(0,0)[bl]{\(\mu_{1}\)}}

\put(18,18){\circle*{1.1}} \put(18,13){\circle*{1.1}}
\put(18,08){\circle*{1.1}}

\put(18,08){\vector(0,1){04}} \put(18,13){\vector(0,1){04}}

\put(23,24.5){\makebox(0,0)[bl]{\(\mu_{2}\)}}

\put(25,23){\circle{2.1}} \put(25,18){\circle{2.1}}
\put(25,13){\circle{2.1}} \put(25,08){\circle{2.1}}

\put(25,23){\circle*{1.0}} \put(25,18){\circle*{1.0}}
\put(25,13){\circle*{1.0}} \put(25,08){\circle*{1.0}}

\put(25,08){\vector(0,1){04}} \put(25,13){\vector(0,1){04}}
\put(25,18){\vector(0,1){04.5}}

\end{picture}
%
\begin{picture}(60,25)
\put(07,00){\makebox(0,0)[bl]{Fig. 12. Balanced sign graph}}

\put(06,07.5){\makebox(0,0)[bl]{``-''}}
\put(18,07.5){\makebox(0,0)[bl]{``+''}}
\put(36,07.5){\makebox(0,0)[bl]{``-''}}
\put(50,07.5){\makebox(0,0)[bl]{``-''}}

\put(03,14.5){\makebox(0,0)[bl]{``+''}}
\put(10.2,13){\makebox(0,0)[bl]{``-''}}

\put(22,14.5){\makebox(0,0)[bl]{``-''}}
\put(33,14.5){\makebox(0,0)[bl]{``-''}}

\put(39,12.5){\makebox(0,0)[bl]{``+''}}

\put(47,13.5){\makebox(0,0)[bl]{``-''}}
\put(54,14.5){\makebox(0,0)[bl]{``+''}}

\put(27,18.5){\makebox(0,0)[bl]{``+''}}
\put(51,18.5){\makebox(0,0)[bl]{``-''}}

\put(00,07){\line(1,0){60}} \put(15,22){\line(1,0){45}}

\put(00,07){\line(1,1){15}}

\put(15,07){\line(0,1){15}}

\put(30,07){\line(-1,1){15}} \put(30,07){\line(1,1){15}}

\put(45,07){\line(0,1){15}}

\put(60,07){\line(-1,1){15}} \put(60,07){\line(0,1){15}}

\put(00,07){\circle{2.0}} \put(00,07){\circle*{1.1}}

\put(15,07){\circle*{1.6}}

\put(30,07){\circle*{1.6}}

\put(45,07){\circle{2.0}} \put(45,07){\circle*{1.1}}

\put(60,07){\circle*{1.6}}


\put(15,22){\circle{2.0}} \put(15,22){\circle*{1.1}}

\put(45,22){\circle{2.0}} \put(45,22){\circle*{1.1}}

\put(60,22){\circle*{1.6}}

\end{picture}
\end{center}

~~

\begin{center}
\begin{picture}(80,41)

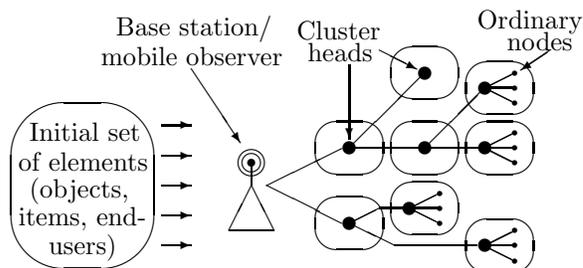
\put(00,00){\makebox(0,0)[bl]{Fig. 13. Hierarchy over balanced
 clusters (network)}}

\put(09.5,16){\oval(19,22)}

\put(2.5,22){\makebox(0,0)[bl]{Initial set}}
\put(1,18){\makebox(0,0)[bl]{of elements}}
\put(02.5,13.5){\makebox(0,0)[bl]{(objects,}}
\put(01,10){\makebox(0,0)[bl]{items, end-}}
\put(05,06){\makebox(0,0)[bl]{users)}}

\put(20,24){\vector(1,0){04}} \put(20,20){\vector(1,0){04}}
\put(20,16){\vector(1,0){04}} \put(20,12){\vector(1,0){04}}
\put(20,08){\vector(1,0){04}}

\put(14,35.1){\makebox(0,0)[bl]{Base station/}}
\put(12,32){\makebox(0,0)[bl]{mobile observer}}

\put(24,31.5){\vector(2,-3){6.5}}

\put(29,10){\line(1,0){6}} \put(29,10){\line(1,2){3}}
\put(35,10){\line(-1,2){3}} \put(32,16){\line(0,1){3}}
\put(32,19){\circle*{1}}

\put(32,19){\circle{2.5}} \put(32,19){\circle{3.5}}

\put(34,16){\line(2,1){10}} \put(34,16){\line(2,-1){10}}

\put(45,21){\oval(09,7)} \put(45,21){\circle*{1.7}}

\put(45,11){\oval(09,7)} \put(45,11){\circle*{1.7}}

\put(45,21){\line(1,0){10}} \put(45,21){\line(1,1){10}}

\put(55,21){\oval(09,7)} \put(55,21){\circle*{1.7}}

\put(55,31){\oval(09,7)} \put(55,31){\circle*{1.7}}

\put(38,35.5){\makebox(0,0)[bl]{Cluster}}
\put(39.5,32.5){\makebox(0,0)[bl]{heads}}

\put(45,32){\vector(0,-1){09.8}} \put(49,34){\vector(2,-1){5}}

\put(55,21){\line(1,0){10}} \put(55,21){\line(1,1){08}}

\put(65,21){\oval(09,7)} \put(63,21){\circle*{1.7}}

\put(63,21){\line(2,1){04}} \put(63,21){\line(1,0){04}}
\put(63,21){\line(2,-1){04}}

\put(67,23){\circle*{0.7}} \put(67,21){\circle*{0.7}}
\put(67,19){\circle*{0.7}}

\put(62,36.2){\makebox(0,0)[bl]{Ordinary}}
\put(66,34){\makebox(0,0)[bl]{nodes}}

\put(71.8,34){\vector(-3,-2){04}}

\put(65,29){\oval(09,7)} \put(63,29){\circle*{1.7}}

\put(63,29){\line(2,1){04}} \put(63,29){\line(1,0){04}}
\put(63,29){\line(2,-1){04}}

\put(67,31){\circle*{0.7}} \put(67,29){\circle*{0.7}}
\put(67,27){\circle*{0.7}}

\put(45,11){\line(2,1){04}} \put(49,13){\line(1,0){06}}
\put(45,11){\line(2,-1){06}} \put(51,8){\line(1,0){11}}

\put(55,13){\oval(09,7)} \put(53,13){\circle*{1.7}}

\put(53,13){\line(2,1){04}} \put(53,13){\line(1,0){04}}
\put(53,13){\line(2,-1){04}}

\put(57,15){\circle*{0.7}} \put(57,13){\circle*{0.7}}
\put(57,11){\circle*{0.7}}

\put(65,8){\oval(09,7)} \put(63,8){\circle*{1.7}}

\put(63,08){\line(2,1){04}} \put(63,08){\line(1,0){04}}
\put(63,08){\line(2,-1){04}}

\put(67,10){\circle*{0.7}} \put(67,8){\circle*{0.7}}
\put(67,6){\circle*{0.7}}

\end{picture}
\end{center}

~~

\begin{center}
\begin{picture}(24,40)
\put(24.5,00){\makebox(0,0)[bl]{Fig. 14. Balanced clustering
 based multi-layer network structure}}

\put(09.5,22){\oval(19,22)}

\put(2.5,26){\makebox(0,0)[bl]{Initial set}}
\put(1,22){\makebox(0,0)[bl]{of elements}}
\put(02.5,17.5){\makebox(0,0)[bl]{(objects,}}
\put(05,14){\makebox(0,0)[bl]{items)}}

\put(20,30){\vector(1,0){04}} \put(20,26){\vector(1,0){04}}
\put(20,22){\vector(1,0){04}} \put(20,18){\vector(1,0){04}}
\put(20,14){\vector(1,0){04}}
\end{picture}
\begin{picture}(116,43)

\put(01,16){\makebox(0,0)[bl]{Domain 2}}

\put(10,10){\oval(20,10)}

\put(01,10){\makebox(0,0)[bl]{\(h^{0}\)}}

\put(02,08){\circle*{1.4}}

\put(02,08){\vector(2,1){4.5}}    \put(07,10.5){\circle*{0.9}}
\put(07,10.5){\vector(2,-1){4.5}} \put(12,08){\circle*{0.9}}
\put(12,08){\vector(1,0){4.5}}

\put(17,08){\circle*{0.8}} \put(17,08){\circle{1.5}}

\put(17,08){\vector(0,1){21}}

\put(36,17){\makebox(0,0)[bl]{Domain 3}}

\put(43,10.5){\oval(26,11)}

\put(32,10){\circle*{0.8}} \put(32,10){\circle{1.5}}

\put(32,10){\vector(4,1){9.5}} \put(42,12.5){\circle*{1.4}}
\put(43.5,12){\makebox(0,0)[bl]{\(h^{g}_{31}\)}}

\put(32,10){\vector(4,-1){9.5}}  \put(42,07.5){\circle*{1.4}}
\put(43.5,05.5){\makebox(0,0)[bl]{\(h^{g}_{33}\)}}

\put(32,10){\vector(1,0){16.5}}  \put(49,10){\circle*{1.4}}
\put(50,08){\makebox(0,0)[bl]{\(h^{g}_{32}\)}}

\put(59,16.5){\makebox(0,0)[bl]{Domain}}
\put(73,17){\makebox(0,0)[bl]{4}}

\put(73,10){\oval(26,10)}

\put(72,10){\circle*{0.8}} \put(72,10){\circle{1.5}}
\put(72,10){\vector(-4,-1){9.5}}    \put(62,07.5){\circle*{1.4}}

\put(61,09){\makebox(0,0)[bl]{\(h^{g}_{41}\)}}

\put(72,10){\vector(1,0){4.5}} \put(77,10){\circle*{0.9}}
\put(77,10){\vector(2,-1){4.5}}    \put(82,07.5){\circle*{1.4}}

\put(80.5,09){\makebox(0,0)[bl]{\(h^{g}_{42}\)}}

\put(98,16){\makebox(0,0)[bl]{Domain 5}}

\put(105,10){\oval(30,10)}

\put(91,06){\makebox(0,0)[bl]{\(h^{g}_{51}\)}}
\put(100.6,06){\makebox(0,0)[bl]{\(h^{g}_{52}\)}}
\put(113,06){\makebox(0,0)[bl]{\(h^{g}_{53}\)}}

\put(92,12.5){\circle{1.5}} \put(92,12.5){\circle*{0.8}}

\put(92,12.5){\vector(1,0){9.5}}    \put(102,12.5){\circle*{0.9}}

\put(102,12.5){\vector(-1,-1){4.5}} \put(97,07.5){\circle*{1.4}}
\put(102,12.5){\vector(1,-1){4.5}} \put(107,07.5){\circle*{1.4}}

\put(102,12.5){\vector(1,0){9.5}}    \put(112,12.5){\circle*{0.9}}
\put(112,12.5){\vector(0,-1){4.5}} \put(112,07.5){\circle*{1.4}}

\put(39,38){\makebox(0,0)[bl]{Domain 1 (Up-layer)}}

\put(55,30){\oval(90,15)}


\put(17,30){\circle*{0.8}} \put(17,30){\circle{1.9}}

\put(17,30){\vector(3,-1){14.5}}

\put(32,25){\circle*{0.8}} \put(32,25){\circle{1.5}}

\put(32,25){\vector(0,-1){14.5}}

\put(17,30){\vector(4,1){19.5}} \put(37,35){\circle*{0.9}}

\put(37,35){\vector(1,0){09.5}} \put(47,35){\circle*{0.9}}

\put(47,35){\vector(2,-1){04.5}} \put(52,32.5){\circle*{0.9}}

\put(52,32.5){\vector(1,0){09.5}} \put(62,32.5){\circle*{0.9}}

\put(52,32.5){\vector(2,-1){09.5}} \put(62,27.5){\circle*{0.9}}

\put(62,27.5){\vector(4,-1){09.5}}

\put(72,25){\circle{1.5}} \put(72,25){\circle*{0.8}}

\put(72,25){\vector(0,-1){14.5}}

\put(62,32.5){\vector(1,0){19.5}} \put(82,32.5){\circle*{0.8}}

\put(82,32.5){\vector(4,-1){09.5}}

\put(92,30){\circle{1.5}} \put(92,30){\circle*{0.8}}

\put(92,30){\vector(0,-1){17}}

\end{picture}
\end{center}

\newpage
\section{Combinatorial balancing problems}

 A list of basic balancing combinatorial optimization
 problems is presented in Table 2.

\begin{center}
 {\bf Table 2.} Basic balancing combinatorial (optimization) problems\\
\begin{tabular}{| c | l | l |}
\hline
 No.&Combinatorial problem &Some sources\\

\hline

 I.&Balanced partitioning problems:&\\

 1.1.&Balanced partition problems (partition of set into balanced subsets)
     &\cite{gar79,hor74,korf98,mert99} \\

 1.2.&\(k\)-balanced partitioning problem
 (partition of set into \(k\) balanced
    &\cite{feld13,feo92}\\

   & subsets)&\\

 1.3.&
 Partitioning (hierarchically clustered) complex networks
    &\cite{lev15c,meyer16a}\\

 &via size-constrained graph clustering&\\

 1.4.&Uniform \(k\)-partition problem&\cite{dell05}\\

 1.5.&Quadratic cost partition problem &\cite{gold05}\\

 1.6.&Multi-constraint partitioning problem
     &\cite{horn06}\\

 1.7.&Tree-balancing problems &\cite{adel62,baer77,karl76,knu97,sle85}\\

 1.8.&Minimum graph bisection problem&\cite{gar79,hrom91}\\

 1.9.&Simple graph partitioning problem&\cite{gar79,soren07}\\

 1.10.& Balanced graph partitioning
 (partition of graph into balanced
  &\cite{andr06,bader13,benlic11,bev14,bul13}\\

 &  components)&\cite{even99,feo90,krau09,macg78}\\

 1.11.&Balanced partitioning of trees &\cite{feld15,lev81,lev15}\\

 1.12.&Balanced partitioning of grids and related graphs
   &\cite{feld12bo}\\

 1.13.&Directed acyclic graph (DAG) partition &\cite{bev16,les09}\\

 1.14.&Graph-clique partitioning problem
      &\cite{bha91,biswas13,char10,erd88,gar79}\\

    &  &\cite{jix07,kim02,patt11,pavan07,rees86}\\

 1.15.&Multiply balanced \(k\)-partitioning of graphs &\cite{amir14}\\

 1.16.&Partitioning a sequence into clusters &\cite{kel16a}\\
    & with restrictions on their cardinalities&\\

 1.17.&General partition data model &\cite{molnar07}\\

\hline
 II.&Some basic balanced combinatorial optimization problems:&\\
 2.1.&Balanced knapsack problems, knapsack load balancing
     &\cite{baldi12,gar79,math98}\\

 2.2.& Balanced bin packing, container loading &\cite{davi99,dell99,gar79}\\

 2.3.&Balanced parallel processor scheduling, balanced job scheduling
     &\cite{changr09,conw67,gar79,saha95,tsail92}\\

   &in grids&\\

 2.4.&Balanced allocation &\cite{azar00,bere00,czu03,kent06}\\

 2.5.&Balanced \(k\)-center problem, balanced \(k\)-weighted center problem
    &\cite{bari93,fel16a,lim05} \\

 2.6.&Matrix balancing&\cite{bali89,cens91,mon97,schn90,zen89}\\

 2.7.&Load balancing problems in distributed systems:&\\

 2.7.1.&Load balancing in computer systems, in multiple processor
       &\cite{chou82,chowy79,cyb89,lev16dafi,lin87}\\

       &  systems&\cite{nil85}\\

 2.7.2.&Load balancing in telecommunications networks&\cite{scho97}\\

 2.7.3.&Load balancing in structuring P2P systems&\cite{rao03}\\

 2.7.4.&Load balancing in web-server systems&\cite{carde99}\\

 2.7.5.&Load balancing in sensor networks &\cite{israr08}\\

 2.7.6.&Special partition of  multi-hop wireless network
       &\cite{gupta10}\\
       &(via graph coloring) for scheduling&\\

 2.8.&Assembly line balancing problems
    &\cite{amen01,andres08,boy07,boy08}\\

   & & \cite{hoff90,saw02}\\

 2.9.&Balanced graph matching &\cite{cour07}\\

 2.10.&Route balancing problems&\cite{joz09,kri10,nik04,oyo14,pand16}\\

 2.11.&Partition and balancing in meshes
      &\cite{diek98,diek00,wal10,walc00,walc01}\\

 2.12.&Balanced combinatorial cooperative games&\cite{fai93}\\

\hline
\end{tabular}
\end{center}

~~

~~

\newpage
\section{Balance/unbalance indices for clustering solution}

 Here, balance/unbalance indices are described for
 clustering solution
 \( \widetilde{X}=\{X_{1},..., X_{\iota},...,X_{\lambda}\}\).
 Let
 \( p^{bal} ( X_{\iota})\) be a total parameter for cluster
 \(X_{\iota}\) (\(\iota = \overline{1,\lambda}\))
 (e.g, the number of elements, total weight).
 In general,
 a balance/unbalance index for clustering solution
   \( \widetilde{X}\)
 can be defined via two methods:

~~

 {\bf Method 1:}
 The balance (unbalance) index is defined as the following:
 \[B(\widetilde{X})=
  \max_{\iota = \overline{1,\lambda}} ~ p^{bal} (X_{\iota})
  -
  \min_{\iota = \overline{1,\lambda}} ~ p^{bal} (X_{\iota}).
 \]
 Note, assessment of \(p^{bal} ( X_{\iota})\)
 can be based on various scales:
 quantitative,
 ordinal,
 poset-like
 (e.g., multiset-based) \cite{lev12a,lev15}.
 Now, the following additional element parameters
 are considered:

 1. The weight of item ~\(w (a_{j}) ~ \forall a_{j} \in A\)
 (e.g., \(w(a_{j} \geq 0 \)).

 2. The weight of edge/arc between items
  ~\(v (a_{j_{1}},a_{j_{2}}) ~  \forall a_{j_{1}},a_{j_{2}} \in  A,
   ~(a_{j_{1}},a_{j_{2}})  \in E \)
  (e.g., \(v (a_{j_{1}},a_{j_{2}})  \geq 0\)).

 3. The structure of cluster by elements types is defined as a special
 multiset estimate \cite{lev12a,lev15}:
 \( e ( X_{\iota} ), \iota = \overline{1,\lambda} \),
 \(e(X_{\iota})=(\theta_{1},...,\theta_{\gamma},...,\theta_{\overline{\gamma}})\),
 where
 \(\theta_{\gamma}\) equals the number of element of type
 \(\gamma\)
  ( \( \gamma=\overline{1,\overline{\gamma} }\) )
  in cluster \(X_{\iota}\),
  the set of element types is:
  \( \{ 1,...,\gamma,..., \overline{\gamma } \} \),
 ~\( \sum_{\gamma = 1}^{\overline{\gamma}}  ~ \theta_{\gamma}
 ~ = ~ | X_{\iota} | \).
 In Fig. 4,
 an example with three element types is depicted
 (clustering solution
 \(\widetilde{X}=\{ X_{1},X_{2},X_{3},X_{4},X_{5},X_{6},X_{7}\}\)).

 In addition, a special ``empty'' element type is considered with
 the number
 \( \overline{\gamma } + 1 \).
 This is a basis to consider
 the clusters of the same size (cardinality)
 to compare their structure:

 for each cluster multiset estimate
 the following element is added:
 \( \theta_{\overline{\gamma} +1}\),
 for cluster of the maximal size
  \( \theta_{\overline{\gamma} +1} = 0\),
 otherwise
  \(\theta_{\overline{\gamma}+1}=| X|^{max}-|X_{\iota}|\),
  where
 \(|X|^{max}\) is the maximal cluster size.

 As a result, the number of components in each multiset estimate of
 cluster structure will be the same.
 The proximity between structure estimates for two clusters
 will be examined as a way (the number of steps)
 from one estimate to another estimate \cite{lev12a,lev15}
 (this is a simplified definition of the multiset estimate
 proximity).
 Note, the element types are linear ordered by
 importance/complexity (example for four types)
 \( 1\succeq 2\succeq 3\succeq 4\).
 The ordering can be considered as a complexity of communication
 node, for example:
  access point (\(1\)),
  transition node (\(2\)),
  end-node (\(3\)),
  ``empty'' element (\(4\)).
 Thus, corresponding lattice based scale can be used
 \cite{lev12a,lev15}.

  As a result,
  the following balance indices for clustering solution can be examined
  (for method 1):

 (1.1) balance index by cluster cardinality is:
 \[ B^{c}( \widetilde{X} ) =
  \max_{\iota = \overline{1,\lambda} }
   ~ | X_{\iota} |
  -
  \min_{ \iota = \overline{1,\lambda} }
   ~ | X_{\iota} |
 .\]
%

 (1.2) balance index by total cluster weight:
 \[ B^{w}( \widetilde{X} ) =
  \max_{\iota = \overline{1,\lambda} }
 ~ \sum_{a_{j}\in X_{\iota}} w (a_{j})
  -
  \min_{ \iota = \overline{1,\lambda} }
 ~ \sum_{a_{j}\in X_{\iota}} w (a_{j})
 .\]
%

 (1.3) balance index by total inter-cluster edge/arc weight:
 \[ B^{v}( \widetilde{X} ) =
  \max_{\iota = \overline{1,\lambda} }
 ~ \sum_{a_{j_{1}},a_{j_{2}}\in X_{\iota}} v (a_{j_{1}},a_{j_{2}}  )
  -
  \min_{\iota = \overline{1,\lambda} }
 ~ \sum_{a_{j_{1}},a_{j_{2}}\in X_{\iota}} v (a_{j_{1}},a_{j_{2}}  )
 .\]
%

 (1.4) balance index by total cluster structure:
 \[ B^{s}( \widetilde{X} ) =
  \max_{\iota = \overline{1,\lambda} }
 ~ e( X_{\iota} )
  -
  \min_{\iota = \overline{1,\lambda} }
 ~ e( X_{\iota} )
 .\]


~~

 {\bf Method 2.} Here, the balance/unbalance index is defined as the
 following:
 \[ \widehat{B}( \widetilde{X} ) =
  \max_{\iota = \overline{1,\lambda} }~~
  |  p^{bal} ( X_{\iota} ) -  p^{0}, \]
 where
 \( p^{0} \) is a special specified
 (e.g., average)  parameter of
 a special reference (e.g., average) cluster
 \(X^{0}\)
 (size,  weight, interconnection weight,
 structure estimate).
 Thus,
  the following balance indices for clustering solution can be examined:

 (2.1) balance index by cluster cardinality is:
 \[ \widehat{B}^{c}( \widetilde{X} ) =
  \max_{\iota = \overline{1,\lambda} }
   ~~ |~~  | X_{\iota} |  -  p_{ |X^{0}| }~~  |,\]
 where \(p_{| X^{0} |}\) is a special specified
 (e.g., average) cluster size.

 (2.2) balance index by total cluster weight:
 \[ \widehat{B}^{w}( \widetilde{X} ) =
  \max_{\iota = \overline{1,\lambda} }
 ~~ | ~ \sum_{a_{j}\in X_{\iota}} w (a_{j})  - p_{w^{0}} ~|
 ,\]
 where \(p_{w^{0}}\) is a special specified
 (e.g., average) cluster weight.

 (2.3) balance index by total inter-cluster edge/arc weight:
 \[ \widehat{B}^{v}( \widetilde{X} ) =
  \max_{\iota = \overline{1,\lambda} }
 ~~ | \sum_{a_{j_{1}},a_{j_{2}}\in X_{\iota}} v (a_{j_{1}},a_{j_{2}}  )
  - p_{v^{0}} ~|
 ,\]
 where \(p_{v^{0}}\) is a special specified
 (e.g., average) cluster weight of inter-cluster interconnections
  (i.e., edges/arcs).

 (2.4) balance index by total cluster structure:
 \[ \widehat{B}^{s}( \widetilde{X} ) =
  \max_{\iota = \overline{1,\lambda} } ~~ | ~ e( X_{\iota}) -
  e_{p^{0}} ~ |
 ,\]
 where \(e_{p^{0}}\) is a special specified
 (e.g., average) multiset estimate of cluster structure.

 Table 3 contains the list of the considered balance indices.
 Note, a hybrid balancing (i.e., by several balance parameters)
 can be examined as well.

~~~

\begin{center}
 {\bf Table 3.} Balance/unbalance indices of clustering solution \(\widetilde{X}\) \\
\begin{tabular}{| c | l | l |}
\hline
 No.&Description &Notation\\

\hline

 I. &Method 1
 (difference between maximal and  &\\

 &minimal values of cluster total parameters): &\\

 1.1.& Balance index by cluster cardinality
     & \( B^{c}( \widetilde{X} )\)\\

 1.2.& Balance index  by total cluster weight
     & \( B^{w}( \widetilde{X} )\)  \\

 1.3.& Balance index  by total inter-cluster
   edge/arc weight
     & \( B^{v}( \widetilde{X} )\)  \\

 1.4.& Balance index  by cluster element structure
     & \( B^{s}( \widetilde{X} )\)  \\

\hline
 II. &Method 2 (maximal difference between  cluster total parameters&\\

 &and special reference specified (fixed) cluster total parameter): &\\

 2.1.& Balance index by cluster cardinality
     & \( \widehat{B}^{c}( \widetilde{X} )\)  \\

 2.2.& Balance index  by total cluster weight
     & \( \widehat{B}^{w}( \widetilde{X} )\)  \\

 2.3.& Balance index  by total inter-cluster
    edge/arc weight
     & \( \widehat{B}^{v}( \widetilde{X} )\)  \\

 2.4.& Balance index  by cluster element structure
     & \( \widehat{B}^{s}( \widetilde{X} )\)  \\

\hline
\end{tabular}
\end{center}

\newpage
\section{Optimization problem formulations for balanced
  clustering}

 Thus, the balanced clustering problem is targeted to search for
 the clustering solution with minimum balance index
 (e.g., index above).
 Clearly,
 additions to the balanced problem statement can involve the
 following:
 (a) some constraints:
 the fixed number of clusters,
 restriction(s) to cluster sizes (including specified integer
 interval), etc.;
 (b) objective function(s)
 (e.g., minimization of total interconnections weight between clusters,
 maximization of total element connections weight in clusters).

 The described
 balanced indices for clustering solutions can be used as
 objective functions and a basis for constraints
 in formulations of optimization balanced clustering models, for
 example:

~~

 {\bf Problem 1:}~

 \[ \min ~ B^{c} ( \widetilde{X} ) ~~~~~
 s.t. ~~  B^{w}(\widetilde{X}) \leq w^{0} , ~~
 B^{s}(\widetilde{X}) \preceq e^{0},\]
 where \(w^{0}\) is a constraint for cluster weight difference,
\(e^{0}\) is a constraint for cluster structure difference.

~~

 {\bf Problem 2:}~

 \[ \min ~ B^{w} ( \widetilde{X} ) ~~~~~
 s.t. ~~   B^{c}(\widetilde{X}) \leq c^{0} , ~~
 B^{s}(\widetilde{X}) \preceq e^{0},\]
 where
  \(c^{0}\) is a  constraint for cluster cardinality difference,
 \(e^{0}\) is a constraint for cluster structure difference.

~~

 {\bf Problem 3:}~

 \[ \min ~ B^{w} ( \widetilde{X} ) ~~~~~
 s.t. ~~  B^{c}(\widetilde{X}) \leq c^{0} , ~~
 |\widetilde{X} | \leq \lambda^{0},\]
 where
  \(c^{0}\) is a  constraint for cluster cardinality difference,
 \(\lambda^{0}\) is a constraint for numbers of clusters
  in the clustering solution.

~~

 Evidently, multicriteria optimization models can be examined as
 well, for example:

~~

 {\bf Problem 4:}~
 \[\min ~ \widetilde{B}^{c} ( \widetilde{X} ), ~~
    \min \widetilde{B}^{v} ( \widetilde{X} )
 ~~~~~~ s.t. ~~    B^{w}(\widetilde{X}) \leq w^{0} , ~~
     \widetilde{B}^{s}(\widetilde{X}) \preceq \widetilde{e}^{0},\]
 where \(w^{0}\) is a constraint for cluster weight difference,
\(\widetilde{e}^{0}\) is a constraint for cluster structure
 difference based on a reference value.

~~

 {\bf Problem 5:}~
 \[\min ~ \widetilde{B}^{c} ( \widetilde{X} ), ~~
   \min \widetilde{B}^{w} ( \widetilde{X} ), ~~
    \min \widetilde{B}^{v} ( \widetilde{X} )
 ~~~~~~ s.t.
  ~~
     \widetilde{B}^{s}(\widetilde{X}) \preceq \widetilde{e}^{0},
  ~~
   |\widetilde{X} | \leq \lambda^{0},\]
 where
 \(\widetilde{e}^{0}\) is a constraint for cluster structure
 difference based on a reference value,
  \(\lambda^{0}\) is a constraint for numbers of clusters
  in the clustering solution.

~~

 {\bf Problem 6:}~
   \[\min B^{w} ( \widetilde{X} ), ~~
    \min \widetilde{B}^{v} ( \widetilde{X} )
 ~~~~~~ s.t. ~~ B^{c}(\widetilde{X}) \leq c^{0} , ~~
     \widetilde{B}^{s}(\widetilde{X}) \preceq \widetilde{e}^{0},\]
 where
  \(c^{0}\) is a  constraint for cluster cardinality difference,
 \(\widetilde{e}^{0}\) is a constraint for cluster structure
 difference based on a reference value.

\newpage
\section{Illustrative examples}

\subsection{Illustration example for description of cluster structure}

 Table 4 contains a description of cluster structures and
 corresponding multiset estimates for example from Fig. 4
 (seven clusters, tree types of elements and additional ``empty'' element type).
 Table 5 contains
 the simplified proximity  (or distance)
 between the multiset estimates of the clusters:~
   \( \delta (e(X_{\iota_{1}}), e(X_{\iota_{2}} ))\),~
   \(\iota_{1},\iota_{2} \in  \{1,...,7\} \).

\begin{center}
 {\bf Table 4.} Cluster structures for example from Fig. 4 \\
\begin{tabular}{|c| c   cccc| c| }
\hline
 Cluster \(X_{\iota}\)  &Element type:&
         \(\theta_{1}\)&\(\theta_{2}\)&\(\theta_{3}\)&\(\theta_{4}\) (``empty'' element)
           & Multiset estimate \(e(X_{\iota})\) \\

\hline

 \(X_{1}\)&&\(1\)&\(1\)&\(3\)&\(2\)&\((1,1,3,2)\)\\

 \(X_{2}\)&&\(1\)&\(1\)&\(3\)&\(2\)&\((1,1,3,2)\)\\

 \(X_{3}\)&&\(1\)&\(1\)&\(4\)&\(1\)&\((1,1,4,1)\)\\

 \(X_{4}\)&&\(1\)&\(1\)&\(3\)&\(2\)&\((1,1,3,2)\)\\

 \(X_{5}\)&&\(1\)&\(1\)&\(2\)&\(3\)&\((1,1,2,3)\)\\

 \(X_{6}\)&&\(2\)&\(1\)&\(4\)&\(0\)&\((2,1,4,0)\)\\

 \(X_{7}\)&&\(1\)&\(2\)&\(4\)&\(0\)&\((1,2,4,0)\)\\

\hline
\end{tabular}
\end{center}

\begin{center}
 {\bf Table 5.} Proximity for cluster structures (example from Fig. 4):
  \( \delta (e(X_{\iota_{1}}), e(X_{\iota_{2}} ))\)
  \\
\begin{tabular}{|c| c   cccc ccc| }
\hline
 Cluster \(X_{\iota_{1}}\) & Cluster \(X_{\iota_{2}}\) :&
         \(X_{1}\)&\(X_{2}\)&\(X_{3}\)&\(X_{4}\)&\(X_{5}\)&\(X_{6}\)&\(X_{7}\)\\

\hline

 \(X_{1}\)&& \(0\)&\(0\)&\(1\)&\(0\)&\(1\)&\(4\)&\(3\)    \\

 \(X_{2}\)&& \(0\)&\(0\)&\(1\)&\(0\)&\(1\)&\(4\)&\(3\)    \\

 \(X_{3}\)&& \(1\)&\(1\)&\(0\)&\(1\)&\(2\)&\(3\)&\(2\)    \\

 \(X_{4}\)&& \(0\)&\(0\)&\(1\)&\(0\)&\(1\)&\(4\)&\(3\)    \\

 \(X_{5}\)&& \(1\)&\(1\)&\(1\)&\(1\)&\(0\)&\(5\)&\(4\)    \\

 \(X_{6}\)&& \(4\)&\(4\)&\(3\)&\(4\)&\(5\)&\(0\)&\(1\)    \\

 \(X_{7}\)&& \(3\)&\(3\)&\(2\)&\(3\)&\(4\)&\(1\)&\(0\)    \\

\hline
\end{tabular}
\end{center}

 For clustering solution from Fig. 4, the following balance
 indices are obtained:

 \( B^{c}( \widetilde{X} ) = 7 - 4 = 3\),
 ~
 \( B^{s}( \widetilde{X} ) =
 e(X_{6}) - e(X_{5}) = (2,1,4,0) - (1,1,2,3) =
 \delta ( e(X_{6}),e(X_{5}) ) = 5\).

 Further, cluster \(X_{1}\) is considered as a reference cluster
 with corresponding its parameters.
 As a result, the following balance indices are obtained:

 \( \widehat{B}^{c}( \widetilde{X} ) = |X_{6}|-|X_{1}|=2\),
 ~
 \( \widehat{B}^{s}( \widetilde{X} ) =
  \delta (e(X_{6}),e(X_{1}))=\delta ((2,1,4,0),(1,1,3,2))=4\).

\subsection{Network-like illustration example}

 In general,
 a wireless sensor network  (WSN)
 can be considered as a multi-layer system
 (Fig. 15)
 \cite{boni07,wangl14}:
 (1) sensors,
 (2) clusters,
 (3) cluster heads,
 (4) sink,
 (5) database server, and
 (6) decision/control center.
 The design process of the architecture is based on
 element multi-type clustering
 (i.e., sensor/end nodes, cluster heads, etc).

\begin{center}
\begin{picture}(116,62)
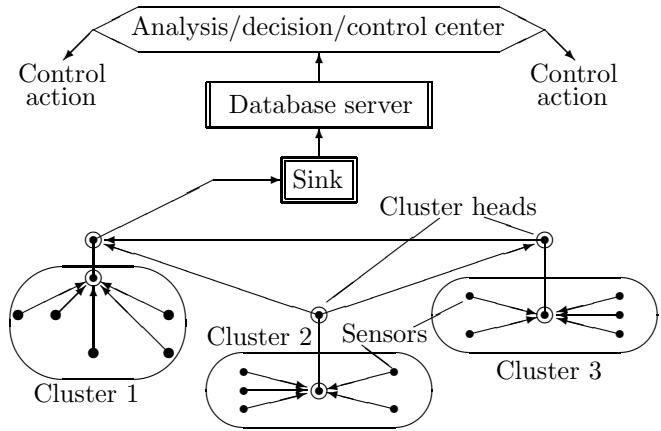
\put(15,00){\makebox(0,0)[bl]{Fig. 15. Multi-layer
 architecture of WSN}}

\put(07,08.5){\makebox(0,0)[bl]{Cluster 1}}

\put(15.6,19){\oval(23.4,15)}

\put(05,20){\circle*{1.4}} \put(25,15){\circle*{1.4}}
\put(15,15){\circle*{1.4}}

\put(25,20){\circle*{1.4}}\put(25,20){\vector(-2,1){09}}

\put(10,20){\circle*{1.4}}\put(10,20){\vector(1,1){04}}

\put(15,25){\circle*{1.0}} \put(15,25){\circle{2.1}}

\put(05,20){\vector(2,1){09}}

\put(25,15){\vector(-1,1){09}} \put(15,15){\vector(0,1){09}}

\put(15,30){\circle*{1.0}} \put(15,30){\circle{2.1}}

\put(15,25){\line(0,1){05}}

\put(30,16){\makebox(0,0)[bl]{Cluster 2}}

\put(45,10){\oval(30,10)}

\put(35,12.5){\circle*{1.2}} \put(55,12.5){\circle*{1.2}}
\put(35,07.5){\circle*{1.2}} \put(55,07.5){\circle*{1.2}}

\put(35,10){\circle*{1.2}} \put(35,10){\vector(1,0){08.4}}

\put(35,12.5){\vector(4,-1){08.4}}
\put(55,12.5){\vector(-4,-1){09}}
\put(35,07.5){\vector(4,1){08.4}} \put(55,07.5){\vector(-4,1){09}}

\put(45,10){\circle*{1.0}} \put(45,10){\circle{2.1}}

\put(45,10){\line(0,1){10}}

\put(45,20){\circle*{1.0}} \put(45,20){\circle{2.1}}

\put(68.5,11.5){\makebox(0,0)[bl]{Cluster 3}}

\put(75,20){\oval(30,10)}

\put(65,22.5){\circle*{1.2}} \put(85,22.5){\circle*{1.2}}

\put(65,17.5){\circle*{1.2}} \put(85,17.5){\circle*{1.2}}

\put(85,20){\circle*{1.2}} \put(85,20){\vector(-1,0){08.4}}

\put(65,22.5){\vector(4,-1){9}} \put(85,22.5){\vector(-4,-1){8.4}}
\put(65,17.5){\vector(4,1){9}} \put(85,17.5){\vector(-4,1){8.4}}

\put(75,20){\circle*{1.0}} \put(75,20){\circle{2.1}}

\put(75,20){\line(0,1){10}}

\put(75,30){\circle*{1.0}} \put(75,30){\circle{2.1}}

\put(75,30){\vector(-1,0){58.6}} \put(45,20){\vector(3,1){28.6}}
\put(45,20){\vector(-3,1){28.6}}


\put(48,16.5){\makebox(0,0)[bl]{Sensors}}
\put(50.7,16){\line(4,-3){04}} \put(57.6,18.5){\line(2,1){06.5}}

\put(53,33.4){\makebox(0,0)[bl]{Cluster~heads}}
\put(61,33){\line(-4,-3){15}} \put(67,33){\line(3,-1){06.5}}

\put(41.5,37){\makebox(0,0)[bl]{Sink}}

\put(40,35){\line(1,0){10}} \put(40,41){\line(1,0){10}}
\put(40,35){\line(0,1){06}} \put(50,35){\line(0,1){06}}

\put(40.5,35.5){\line(1,0){09}} \put(40.5,40.5){\line(1,0){09}}
\put(40.5,35.5){\line(0,1){05}} \put(49.5,35.5){\line(0,1){05}}

\put(15,30){\line(2,1){16}} \put(31,38){\vector(1,0){09}}

\put(33,47){\makebox(0,0)[bl]{Database server}}

\put(30,45){\line(1,0){30}} \put(30,51){\line(1,0){30}}
\put(30,45){\line(0,1){06}} \put(60,45){\line(0,1){06}}
\put(30.5,45){\line(0,1){06}} \put(59.5,45){\line(0,1){06}}

\put(45,41){\vector(0,1){04}}

\put(20,56.4){\makebox(0,0)[bl]{Analysis/decision/control center}}

\put(21,55){\line(1,0){48}} \put(21,61){\line(1,0){48}}
\put(45,51){\vector(0,1){04}}

\put(21,55){\line(-2,1){6}} \put(21,61){\line(-2,-1){6}}
\put(69,55){\line(2,1){6}} \put(69,61){\line(2,-1){6}}

\put(05,51){\makebox(0,0)[bl]{Control}}
\put(06,48){\makebox(0,0)[bl]{action}}

\put(15,58){\vector(-1,-1){04}} \put(75,58){\vector(1,-1){04}}

\put(73,51){\makebox(0,0)[bl]{Control}}
\put(74,48){\makebox(0,0)[bl]{action}} \

\end{picture}
\end{center}

 Here, a numerical example from Fig. 10 is considered
 (three types of elements:
 ordinary nodes, nodes  for relay,
 cluster heads):
 \(A = \{1,2,3,4,5,6,7,8,9,10,11,12,13,14,15\}\).
 Parameters of elements and their interconnection are presented in
 Table 6 (element weights, element types) and
 Table 7 (weights of edges, symmetric weighted binary relation).
 The considered four-cluster solution is:
 \(\widetilde{X}' = \{ X_{1}', X_{2}', X_{3}', X_{4}'\} \)
 where
 \(X_{1}' = \{8,9,13,14\} \).
 \(X_{2}' = \{1,3,4\} \),
 \(X_{3}' = \{2,5,6,10\} \), and
 \(X_{4}' = \{7,11,12,15\} \).
 Proximities between cluster structures are presented in Table 8.
 Fig. 16 depicts the poset-like scale \(P^{4,4}\)
  \cite{lev12a,lev15}
 for cluster structure, where multiset estimate of the cluster structure is:
 ~\(e( X_{\iota} ) =
 (\alpha_{\theta_{1}},\alpha_{\theta_{2}},\alpha_{\theta_{3}},\alpha_{\theta_{4}})\).

\begin{center}
\begin{picture}(116,154)
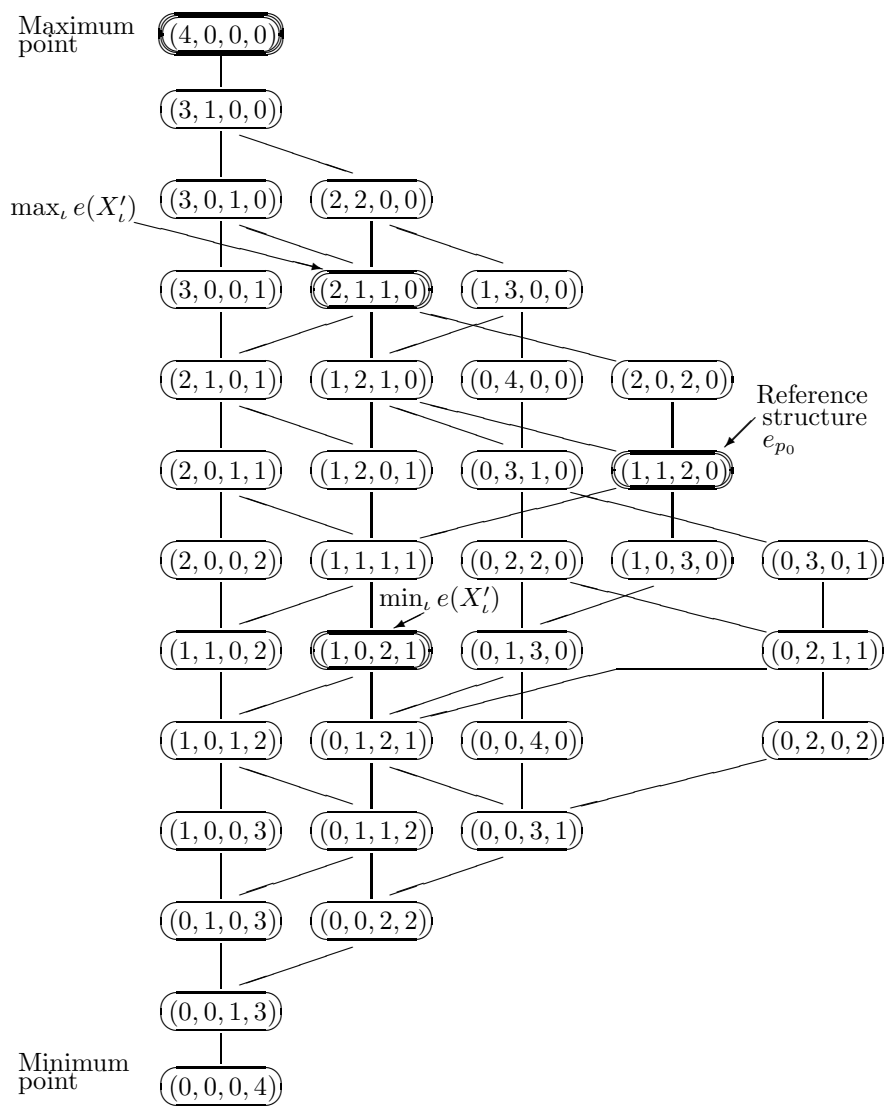
\put(02.5,00){\makebox(0,0)[bl] {Fig. 16. Poset-like scale
 \(P^{4,4}\) for
 ~\(e(X_{\iota}')=(\alpha_{\theta_{1}},\alpha_{\theta_{2}},\alpha_{\theta_{3}},\alpha_{\theta_{4}})\)
 \cite{lev12a,lev15}
 }}

\put(01,149){\makebox(0,0)[bl]{Maximum}}
\put(01,146){\makebox(0,0)[bl]{point}}

\put(21,147){\makebox(0,0)[bl]{\((4,0,0,0)\) }}
\put(28,149){\oval(16,5)} \put(28,149){\oval(16.5,5.5)}
\put(28,149){\oval(15.5,4.5)}


\put(28,142){\line(0,1){4}}

\put(21,137){\makebox(0,0)[bl]{\((3,1,0,0)\)}}
\put(28,139){\oval(16,5)}

\put(28,130){\line(0,1){6}}

\put(21,125){\makebox(0,0)[bl]{\((3,0,1,0)\) }}
\put(28,127){\oval(16,5)}


\put(00,124){\makebox(0,0)[bl]{\(\max_{\iota} e(X_{\iota}')\)}}
\put(16.5,124){\vector(4,-1){25}}


\put(28,118){\line(0,1){6}}
\put(21,113){\makebox(0,0)[bl]{\((3,0,0,1)\)}}
\put(28,115){\oval(16,5)}

\put(28,106){\line(0,1){6}}

\put(21,101){\makebox(0,0)[bl]{\((2,1,0,1)\) }}
\put(28,103){\oval(16,5)}


\put(28,94){\line(0,1){6}}

\put(21,89){\makebox(0,0)[bl]{\((2,0,1,1)\)}}
\put(28,91){\oval(16,5)}

\put(28,82){\line(0,1){6}}

\put(21,77){\makebox(0,0)[bl]{\((2,0,0,2)\) }}
\put(28,79){\oval(16,5)}


\put(28,70){\line(0,1){6}}
\put(21,65){\makebox(0,0)[bl]{\((1,1,0,2)\)}}
\put(28,67){\oval(16,5)}

\put(28,58){\line(0,1){6}}
\put(21,53){\makebox(0,0)[bl]{\((1,0,1,2)\) }}
\put(28,55){\oval(16,5)}

\put(28,46){\line(0,1){6}}
\put(21,41){\makebox(0,0)[bl]{\((1,0,0,3)\) }}
\put(28,43){\oval(16,5)}

\put(28,34){\line(0,1){6}}
\put(21,29){\makebox(0,0)[bl]{\((0,1,0,3)\) }}
\put(28,31){\oval(16,5)}

\put(28,22){\line(0,1){6}}
\put(21,17){\makebox(0,0)[bl]{\((0,0,1,3)\) }}
\put(28,19){\oval(16,5)}

\put(28,12){\line(0,1){4}}

\put(21,07){\makebox(0,0)[bl]{\((0,0,0,4)\) }}
\put(28,09){\oval(16,5)}

\put(01,11){\makebox(0,0)[bl]{Minimum}}
\put(01,08){\makebox(0,0)[bl]{point}}


\put(30.5,135.5){\line(3,-1){15}}

\put(41,125){\makebox(0,0)[bl]{\((2,2,0,0)\) }}
\put(48,127){\oval(16,5)}

\put(45.5,118.5){\line(-3,1){15}}

\put(45.5,111.5){\line(-3,-1){15}}

\put(48,118){\line(0,1){6}}
\put(41,113){\makebox(0,0)[bl]{\((2,1,1,0)\) }}
\put(48,115){\oval(16,5)} \put(48,115){\oval(15.4,4.4)}

\put(48,106){\line(0,1){6}}
\put(41,101){\makebox(0,0)[bl]{\((1,2,1,0)\) }}
\put(48,103){\oval(16,5)}

\put(45.5,94.5){\line(-3,1){15}}

\put(48,94){\line(0,1){6}}

\put(41,89){\makebox(0,0)[bl]{\((1,2,0,1)\) }}
\put(48,91){\oval(16,5)}

\put(45.5,82.5){\line(-3,1){15}}

\put(45.5,75.5){\line(-3,-1){15}}
\put(48,82){\line(0,1){6}}

\put(41,77){\makebox(0,0)[bl]{\((1,1,1,1)\) }}
\put(48,79){\oval(16,5)}
\put(45.5,63.5){\line(-3,-1){15}}

\put(48,70){\line(0,1){6}}
\put(41,65){\makebox(0,0)[bl]{\((1,0,2,1)\) }}
\put(48,67){\oval(16,5)} \put(48,67){\oval(15.4,4.4)}

\put(49,72){\makebox(0,0)[bl]{\(\min_{\iota} e(X_{\iota}')\)}}
\put(55,72){\vector(-2,-1){04}}

\put(48,58){\line(0,1){6}}
\put(41,53){\makebox(0,0)[bl]{\((0,1,2,1)\) }}
\put(48,55){\oval(16,5)}
\put(45.5,46.5){\line(-3,1){15}}

\put(45.5,39.5){\line(-3,-1){15}}

\put(48,46){\line(0,1){6}}

\put(41,41){\makebox(0,0)[bl]{\((0,1,1,2)\) }}
\put(48,43){\oval(16,5)}
\put(45.5,27.5){\line(-3,-1){15}}

\put(48,34){\line(0,1){6}}
\put(41,29){\makebox(0,0)[bl]{\((0,0,2,2)\) }}
\put(48,31){\oval(16,5)}

\put(65.5,118.5){\line(-3,1){15}}
\put(65.5,111.5){\line(-3,-1){15}}

\put(61,113){\makebox(0,0)[bl]{\((1,3,0,0)\) }}
\put(68,115){\oval(16,5)}

\put(68,106){\line(0,1){6}}

\put(61,101){\makebox(0,0)[bl]{\((0,4,0,0)\) }}
\put(68,103){\oval(16,5)}

\put(65.5,94.5){\line(-3,1){15}}

\put(68,94){\line(0,1){6}}

\put(61,89){\makebox(0,0)[bl]{\((0,3,1,0)\) }}
\put(68,91){\oval(16,5)}

\put(68,82){\line(0,1){6}}

\put(61,77){\makebox(0,0)[bl]{\((0,2,2,0)\) }}
\put(68,79){\oval(16,5)}

\put(65.5,63.5){\line(-3,-1){15}}

\put(68,70){\line(0,1){6}}

\put(61,65){\makebox(0,0)[bl]{\((0,1,3,0)\) }}
\put(68,67){\oval(16,5)}

\put(68,58){\line(0,1){6}}

\put(61,53){\makebox(0,0)[bl]{\((0,0,4,0)\) }}
\put(68,55){\oval(16,5)}

\put(65.5,46.5){\line(-3,1){15}} \put(65.5,39.5){\line(-3,-1){15}}

\put(68,46){\line(0,1){6}}

\put(61,41){\makebox(0,0)[bl]{\((0,0,3,1)\) }}
\put(68,43){\oval(16,5)}

\put(80.5,105.5){\line(-4,1){26}}

\put(81,101){\makebox(0,0)[bl]{\((2,0,2,0)\) }}
\put(88,103){\oval(16,5)}

\put(80.5,93.5){\line(-4,1){26}} \put(80.5,88.5){\line(-4,-1){26}}

\put(88,94){\line(0,1){6}}

\put(81,89){\makebox(0,0)[bl]{\((1,1,2,0)\) }}
\put(88,91){\oval(16,5)} \put(88,91){\oval(15.4,4.4)}

\put(85.5,75.5){\line(-3,-1){15}}

\put(88,82){\line(0,1){6}}

\put(81,77){\makebox(0,0)[bl]{\((1,0,3,0)\) }}
\put(88,79){\oval(16,5)}

\put(100.5,81.5){\line(-4,1){26}}

\put(101,77){\makebox(0,0)[bl]{\((0,3,0,1)\) }}
\put(108,79){\oval(16,5)}

\put(108,70){\line(0,1){6}}

\put(100.5,69.5){\line(-4,1){26}}

\put(100.5,64.5){\line(-1,0){20}}

\put(80.5,64.5){\line(-4,-1){26}}

\put(101,65){\makebox(0,0)[bl]{\((0,2,1,1)\) }}
\put(108,67){\oval(16,5)}

\put(108,58){\line(0,1){6}}

\put(101,53){\makebox(0,0)[bl]{\((0,2,0,2)\) }}
\put(108,55){\oval(16,5)}

\put(100.5,52.5){\line(-4,-1){26}}

\put(99.2,100){\makebox(0,0)[bl]{Reference}}
\put(100,97){\makebox(0,0)[bl]{structure}}
\put(100,93){\makebox(0,0)[bl]{\(e_{p_{0}}\)}}

\put(99,98){\vector(-1,-1){4}}

\end{picture}
\end{center}

 The resultant integrated parameters of the clusters are contained
 in Table 9.

 Finally, the following balance indices on the basis of method 1
 are obtained (clustering solution
 \(\widetilde{X}' = \{ X_{1}', X_{2}',X_{3}',X_{4}' \} \)):~
 \( B^{c}( \widetilde{X}') = 1\),
 \( B^{w}( \widetilde{X}' ) = 6.7\),
 \( B^{v}( \widetilde{X}' ) = 13.9\),
 \( B^{s}( \widetilde{X}' ) =
   4  \).

 Further, the following reference specified parameters for clusters
 are considered:
 \(p_{|X^{0}|} = 4\),
 \(p_{w^{0}} = 12.0\),
 \(p_{v^{0}} = 15.0\).

 The reference cluster structure is:
 \(e_{p^{0}} = (1,1,2,0)\).
 Thus, the balance indices based on method 2 are:
 \(\widehat{B}^{c}( \widetilde{X}')=1\),
 \(\widehat{B}^{w}( \widetilde{X}')=4.7\),
 \(\widehat{B}^{v}( \widetilde{X}')=7.3\),
 \(\widehat{B}^{s}( \widetilde{X}')=2\).

\newpage
\begin{center}
 {\bf Table 6.} Parameters of elements (example from Fig. 10) \\
\begin{tabular}{| c | c | c |c|}
\hline
 Element \(a_{j}\)&Cluster number \(\iota\) (\(X_{\iota}'\)  )
 &Element weight \(w(a_{j})\) &Element type \(\theta(a_{j})\)\\

\hline

 \(a_{1}\)&\(2\)& \(4.2\)  & \(1\)  \\

 \(a_{2}\)&\(3\)& \(5.1\)  & \(1\)  \\

 \(a_{3}\)&\(2\)& \(1.1\)  & \(3\)  \\

 \(a_{4}\)&\(2\)& \(2.0\)  & \(3\)  \\

 \(a_{5}\)&\(3\)& \(3.1\)  & \(2\)  \\

 \(a_{6}\)&\(3\)& \(3.2\)  & \(2\)  \\

 \(a_{7}\)&\(4\)& \(1.0\)  & \(3\)  \\

 \(a_{8}\)&\(1\)& \(3.4\)  & \(2\)  \\

 \(a_{9}\)&\(1\)& \(5.0\)  & \(1\)  \\

 \(a_{10}\)&\(3\)& \(0.9\)  & \(3\)  \\

 \(a_{11}\)&\(4\)& \(4.5\)  & \(1\)  \\

 \(a_{12}\)&\(4\)& \(4.8\)  & \(1\)  \\

 \(a_{13}\)&\(1\)& \(0.8\)  & \(3\)  \\

 \(a_{14}\)&\(1\)& \(3.4\)  & \(2\)  \\

 \(a_{15}\)&\(4\)& \(3.7\)  & \(2\)  \\

\hline
\end{tabular}
\end{center}

\begin{center}
 {\bf Table 7.} Interconnection edge weights for example from Fig. 10 \\
\begin{tabular}{|c| c   ccccc   ccccc  ccccc| }
\hline
 \(a_{j_{1}}\)&\(a_{j_{2}}\):&
         \(1\)& \(2\)&\(3\)&\(4\)&\(5\)&
         \(6\)&\(7\)&\(8\)&\(9\)&\(10\)&
         \(11\)&\(12\)&\(13\)&\(14\)&\(15\)
         \\
\hline

 \(1\)&&
        \(\)& \(\)&\(4.1\)&\(2.1\)&\(\)&
        \(\)&\(\)&\(\)&\(\)&\(\)&
        \(\)&\(\)&\(\)&\(\)&\(\) \\

 \(2\)&&
        \( \)&\( \)&\(\)&\(\)&\(4.4\)&
        \(4.5\)&\(\)&\(\)&\(\)&\(\)&
        \(\)&\(\)&\(\)&\(\)&\(\)  \\

 \(3\)&&
        \(4.1\)&\( \)&\( \)&\(1.5\)&\(\)&
        \(\)&\(\)&\(\)&\(\)&\(\)&
        \(\)&\(\)&\(\)&\(\)&\(\)  \\

 \(4\)&&
        \(2.1\)&\( \)&\(1.5\)&\( \)&\(\)&
        \(\)&\(\)&\(3.0\)&\(2.9\)&\(0.7\)&
        \(\)&\(\)&\(\)&\(\)&\(\)  \\

 \(5\)&&
        \( \)& \(4.4\)&\( \)&\( \)&\( \)&
        \(3.6\)&\(\)&\(\)&\(3.0\)&\(1.0\)&
        \(\)&\(\)&\(\)&\(\)&\(\) \\

 \(6\)&&
        \( \)& \(4.5\)&\( \)&\( \)&\(3.6\)&
        \( \)&\(\)&\(\)&\(\)&\(0.8\)&
        \(\)&\(\)&\(\)&\(\)&\(\) \\

 \(7\)&&
        \( \)& \( \)&\( \)&\( \)&\( \)&
        \( \)&\( \)&\(\)&\(\)&\(\)&
        \(2.5\)&\(3.2\)&\(\)&\(\)&\(\) \\

 \(8\)&&
        \( \)&\( \)&\( \)&\(3.0\)&\( \)&
        \( \)&\( \)&\( \)&\(4.0\)&\(\)&
        \(\)&\(\)&\(\)&\(3.2\)&\(\) \\

 \(9\)&&
        \( \)& \( \)&\( \)&\(2.9\)&\(3.0\)&
        \( \)&\( \)&\(4.0\)&\( \)&\(1.6\)&
        \( \)&\( \)&\(3.1\)&\(6.0\)&\( \) \\

 \(10\)&&
        \( \)&\( \)&\( \)&\(0.7\)&\(1.0\)&
        \(0.8\)&\( \)&\( \)&\(1.6\)&\( \)&
        \(5.0\)&\(\)&\(\)&\(\)&\(3.3\)  \\

 \(11\)&&
        \( \)& \( \)&\( \)&\( \)&\( \)&
        \( \)&\(2.5\)&\( \)&\( \)&\(5.0\)&
        \( \)&\(6.2\)&\(\)&\(\)&\(4.3\) \\

 \(12\)&&
        \( \)&\( \)&\( \)&\( \)&\( \)&
        \( \)&\(3.2\)&\( \)&\( \)&\( \)&
        \(6.2\)&\( \)&\( \)&\( \)&\(4.2\) \\

 \(13\)&&
        \( \)&\( \)&\( \)&\( \)&\( \)&
        \( \)&\( \)&\( \)&\(3.1\)&\( \)&
        \( \)&\( \)&\( \)&\(5.0\)&\( \) \\

 \(14\)&&
        \( \)&\( \)&\( \)&\( \)&\( \)&
        \( \)&\( \)&\(3.2\)&\(6.0\)&\( \)&
        \( \)&\( \)&\(5.0\)&\( \)&\(2.5\)  \\

 \(15\)&&
        \( \)&\( \)&\( \)&\( \)&\( \)&
        \( \)&\( \)&\( \)&\( \)&\(3.3\)&
        \(4.3\)&\(4.2\)&\( \)&\(2.5\)&\(\)  \\

\hline
\end{tabular}
\end{center}

\begin{center}
 {\bf Table 8.} Proximity for cluster structures (example from Fig. 10):
  \( \delta (e(X_{\iota_{1}}'), e(X_{\iota_{2}}' ))\)
  \\
\begin{tabular}{|c| c   cccc| }
\hline
 Cluster \(X_{\iota_{1}}'\) & Cluster \(X_{\iota_{2}}'\) :&
         \(X_{1}'\)&\(X_{2}'\)&\(X_{3}'\)&\(X_{4}'\)
           \\

\hline

 \(X_{1}'\)&& \(0\)&\(2\)&\(0\) &\(1\)  \\

 \(X_{2}'\)&& \(2\)&\(0\)&\(3\) &\(4\)  \\

 \(X_{3}'\)&& \(0\)&\(3\)&\(0\) &\(1\)  \\

 \(X_{4}'\)&& \(1\)&\(4\)&\(1\) &\(0\)  \\

\hline
\end{tabular}
\end{center}

\begin{center}
 {\bf Table 9.} Parameters of clusters (example from Fig. 10) \\
 \begin{tabular}{|c| c | c | c | c | }
\hline
  Cluster& \( \eta_{\iota} ( X_{\iota}' ) = | X_{\iota}' |\)&
           \( \sum_{a_{j}\in X_{\iota}'} w (a_{j}) \) &
           \( \sum_{a_{j_{1}},a_{j_{2}}\in X_{\iota}'} v (a_{j_{1}},a_{j_{2}}) \)&
           \( e( X_{\iota}') \)  \\

\hline

 \( X_{1}'\)& \(4\) & \(12.6\) & \(21.3\)& \((1,2,1,0)\)\\

 \( X_{2}'\)& \(3\) & \(7.3\)  & \(7.7\) & \((1,0,2,1)\)\\

 \( X_{3}'\)& \(4\) & \(12.3\) & \(15.4\)& \((1,2,1,0)\)\\

 \( X_{4}'\)& \(4\) & \(14.0\) & \(21.6\)& \((2,1,1,0)\)\\

\hline
\end{tabular}
\end{center}

%
 The second considered four-cluster solution is:
 \(\widetilde{X}'' = \{ X_{1}'', X_{2}'', X_{3}'', X_{4}''\} \)
 where
 \(X_{1}'' = \{4,8,9,13,14\} \).
 \(X_{2}'' = \{1,3\} \),
 \(X_{3}'' = \{2,5,6\} \), and
 \(X_{4}'' = \{7,10,11,12,15\} \).

 Proximities between cluster structures are presented in Table 10.
 The resultant integrated parameters of the clusters are contained in Table 11.

 Finally, the following balance indices on the basis of method 1 are obtained (clustering solution
 \(\widetilde{X}'' = \{ X_{1}'', X_{2}'',X_{3}'',X_{4}'' \} \)):~
 \( B^{c}( \widetilde{X}'') = 3\),
 \( B^{w}( \widetilde{X}'' ) = 9.6\),
 \( B^{v}( \widetilde{X}'' ) = 25.8\),
 \( B^{s}( \widetilde{X}'' ) = 6  \).

 The basic reference specified parameters for clusters are considered
 as for previous clustering solution (i.e., \(\widetilde{X}'\)):
 \(p_{|X^{0}|} = 4\),
 \(p_{w^{0}} = 12.0\),
 \(p_{v^{0}} = 15.0\).

 The reference cluster structure is:
 \(e_{p^{0}} = (1,1,3,0)\),
 Thus, the balance indices based on method 2 are:
 \( \widehat{B}^{c}( \widetilde{X}'') = 2\),
 \( \widehat{B}^{w}( \widetilde{X}'' ) = 6.7\),
 \( \widehat{B}^{v}( \widetilde{X}'' ) = 11.9\),
 \( \widehat{B}^{s}( \widetilde{X}'' ) = 4\).


 Table 12 contains the obtained balance indices for the considered
 clustering solutions \(\widetilde{X}'\) and \(\widetilde{X}''\).
%

\begin{center}
 {\bf Table 10.} Proximity for cluster structures (example from Fig. 10):
  \( \delta (e(X_{\iota_{1}}''), e(X_{\iota_{2}}'' ))\)
  \\
\begin{tabular}{|c| c   cccc| }
\hline
 Cluster \(X_{\iota_{1}}''\) & Cluster \(X_{\iota_{2}}''\) :&
         \(X_{1}''\)&\(X_{2}''\)&\(X_{3}''\)&\(X_{4}''\)
           \\

\hline
 \(X_{1}''\)&& \(0\)&\(5\)&\(2\) &\(1\) \\

 \(X_{2}''\)&& \(5\)&\(0\)&\(3\) &\(6\) \\

 \(X_{3}''\)&& \(2\)&\(3\)&\(0\) &\(3\) \\

 \(X_{4}''\)&& \(1\)&\(6\)&\(3\) &\(0\) \\

\hline
\end{tabular}
\end{center}

\begin{center}
 {\bf Table 11.} Parameters of clusters (example from Fig. 10) \\
 \begin{tabular}{|c| c | c | c | c | }
\hline
  Cluster& \( \eta_{\iota} ( X_{\iota}'' ) =  | X_{\iota}'' |\)&
           \( \sum_{a_{j}\in X_{\iota}''} w (a_{j}) \) &
           \( \sum_{a_{j_{1}},a_{j_{2}}\in X_{\iota}''} v (a_{j_{1}},a_{j_{2}}) \)&
           \( e( X_{\iota}'') \)  \\

\hline

 \( X_{1}''\)& \(5\) & \(14.6\) & \(27.2\)& \((1,2,2,0)\)\\

 \( X_{2}''\)& \(2\) & \(5.3\)  & \(4.1\) & \((1,0,1,3)\)\\

 \( X_{3}''\)& \(3\) & \(11.4\) & \(12.5\)& \((1,2,0,2)\)\\

 \( X_{4}''\)& \(5\) & \(14.9\) & \(29.9\)& \((2,1,2,0)\)\\

\hline
\end{tabular}
\end{center}

\begin{center}
 {\bf Table 12.} Balance indices of clustering solutions
 \(\widetilde{X}'\) and \(\widetilde{X}''\)\\
\begin{tabular}{| c | l | c | c | }
\hline
 No.&Description&Solution \(\widetilde{X}'\)&Solution \(\widetilde{X}''\)\\

\hline

 I. &Method 1:&&\\

 1.1.&Balance index by cluster cardinality
     &\(B^{c}(\widetilde{X}')=1\)
     &\(B^{c}(\widetilde{X}'')=3\) \\

 1.2.&Balance index by total cluster weight
     &\(B^{w}(\widetilde{X}')=6.7\)
     &\(B^{w}(\widetilde{X}'')=9.6\) \\

 1.3.& Balance index  by total inter-cluster
   edge/arc weight
     &\(B^{v}(\widetilde{X}')=13.9\)
     &\(B^{v}(\widetilde{X}'')=25.8\) \\

 1.4.& Balance index  by cluster element structure
     &\( B^{s}(\widetilde{X}')=4\)
     &\( B^{s}(\widetilde{X}'')=6\) \\

\hline
 II. &Method 2: &&\\

 2.1.&Balance index by cluster cardinality
     &\(\widehat{B}^{c}(\widetilde{X}')=1\)
     &\(\widehat{B}^{c}(\widetilde{X}'')=2\) \\

 2.2.&Balance index  by total cluster weight
     &\(\widehat{B}^{w}(\widetilde{X}')=4.7\)
     &\(\widehat{B}^{w}(\widetilde{X}'')=6.7\) \\

 2.3.&Balance index  by total inter-cluster
    edge/arc weight
     &\(\widehat{B}^{v}(\widetilde{X}')=7.3\)
     &\(\widehat{B}^{v}(\widetilde{X}'')=11.9\) \\

 2.4.& Balance index  by cluster element structure
     &\(\widehat{B}^{s}(\widetilde{X}')=2\)
     &\(\widehat{B}^{s}(\widetilde{X}'')=4\)  \\

\hline
\end{tabular}
\end{center}

\subsection{Cluster structure based balanced
 clustering for student teams}

 The considered numerical example is a modification
  of an example from \cite{lev15c}.
 There is a set of \(13\) students
 \(A = \{a_{1},...,a_{j},...,a_{13}\}\)
  (Table 13)
  in the field of radio engineering.
 Four inclination/skill properties of the students are considered as
 parameters/criteria (student estimates are shown in Table 13):
 (i) inclination (skill) for mathematics \(C_{1}\),
 (2) inclination  (skill) for theoretical radio engineering  \(C_{2}\),
 (3) skill for technical works in radio engineering
 (usage of radio devices, design of scheme, analysis of signals, etc.)
 \(C_{3}\),
 (4) writing skill (e.g., to prepare a laboratory work report/paper)
    \(C_{4}\).
 The following  scale is used:
 \([0,1,2,3]\), where
 \(0\) corresponds to absent inclination/skill),
 \(1\) corresponds to low level of inclination/skill),
 \(2\) corresponds to medium level of inclination/skill),
 \(3\) corresponds to high level of inclination/skill).
 As a result, each student \(j\) (\(j=\overline{1,13}\)) has a vector (4 component)
 estimate~
 \(\xi(a_{j}) = (\xi^{1}(a_{j}), \xi^{2}(a_{j}), \xi^{3}(a_{j}), \xi^{4}(a_{j}) ) \).
  It is assumed,
  each student has as minimum one ``positive'' inclination/skill estimate (or more).

 A symmetric weighted binary relation of student friendship
 \(R^{f} = \{ \varepsilon (a_{j_{1}},a_{j_{2}}) \}\)
 (\(j_{1},j_{2} = \overline{1,13}\))
 (ordinal scale \([0,1,2,3]\), \(0\) corresponds to
 incompatibility) is contained in Table 14.

 Some notations are as follows.
 The structure of  cluster is
 \(X_{\iota}=\{b_{1},...,b_{\tau},...,b_{\mu_{\iota}}\}\).

 Then the following cluster characteristics are examined:

 1. The total vector estimate of the cluster structure is:~

 \(
 \xi (X_{\iota}) =
 (
 \xi^{1}(X_{\iota}),\xi^{2}(X_{\iota}),\xi^{3}(X_{\iota}),\xi^{4}(X_{\iota})
 )
 \),

 where~
 \( \xi^{\kappa} ( X_{\iota} ) =
  \max_{\tau = \overline{1,\mu_{\iota}}}~   \xi^{\kappa} (b_{\tau}) \),
 ~\(\forall \kappa = \overline{1,4}\) ~(index of criteria/parameter).

 2.  The total estimate of the cluster by quality of intercluster
 compatibility is:

 \( \varepsilon (X_{\iota}) =
 \sum_{ b_{\tau_{1}},b_{\tau_{2}} \in  X_{\iota}, b_{\tau_{1}} < b_{\tau_{2}}  } ~\varepsilon (b_{\tau_{1}},b_{\tau_{2}})
 \).

\begin{center}
 {\bf Table 13.} Items/students,  estimates upon criteria   \\
\begin{tabular}{| c |  c   c   c   c  | c|}
\hline
 Item (student) &\(C_{1}\) &\(C_{2}\) &\(C_{3}\) &\(C_{4}\) &Vector estimate    \\
  \(a_{j}\)&(\(\xi^{1}(a_{j})\))&(\(\xi^{2}(a_{j})\))&(\(\xi^{3}(a_{j})\))&(\(\xi^{4}(a_{j})\))&
   \(\xi(a_{j}) \)\\
%
\hline
 Student 1 (\(a_{1}\))& 1 & 2 & 3 & 3 &(1,2,3,3)\\

 Student 2  (\(a_{2}\))& 0 & 1 & 2 & 1 &(0,1,2,1)\\

 Student 3  (\(a_{3}\))& 2 & 3 & 3 & 2 &(1,3,3,2)\\

 Student 4  (\(a_{4}\))& 2 & 2 & 1 & 3 &(3,2,1,3)\\

 Student 5  (\(a_{5}\))& 0 & 1 & 2 & 1 &(0,1,2,1)\\

 Student 6  (\(a_{6}\))& 3 & 3 & 3 & 3 &(3,3,3,3)\\

 Student 7  (\(a_{7}\))& 0 & 1 & 1 & 1 &(0,1,1,1)\\

 Student 8  (\(a_{8}\))& 0 & 2 & 2 & 2 &(0,2,2,2)\\

 Student 9  (\(a_{9}\))& 3 & 3 & 3 & 3 &(3,3,3,3)\\

 Student 10 (\(a_{10}\))& 3 & 3 & 2 & 3 &(3,3,2,3)\\

 Student 11 (\(a_{11}\))& 0 & 1 & 3 & 2 &(0,1,3,2)\\

 Student 12 (\(a_{12}\))& 0 & 2 & 3 & 1 &(0,2,3,1)\\

 Student 13 (\(a_{13}\))& 0 & 1 & 1 & 1 &(0,1,1,1)\\

\hline
\end{tabular}
\end{center}

\begin{center}
 {\bf Table 14.} Ordinal estimates of student friendship
 (compatibility)
  \(R^{f} = \{ \varepsilon (a_{j_{1}},a_{j_{2}}) \}\)
  \\
\begin{tabular}{  |c|c|c|c| c|c|c|c |c|c|c|c | c | }
\hline
    \(a_{j_{1}}\) /  \(a_{j_{2}}\)
  &\(a_{2}\) &\(a_{3}\) &\(a_{4}\) &\(a_{5}\) &\(a_{6}\) &\(a_{7}\) &\(a_{8}\)
  &\(a_{9}\)&\(a_{10}\)&\(a_{11}\)&\(a_{12}\)&\(a_{13}\) \\
\hline

 \(a_{1}\) &\(2\)&\(3\)&\(3\)&\(3\)&\(3\)&\(1\)&\(2\)&\(2\)&\(2\)&\(2\)&\(3\)&\(2\)  \\

 \(a_{2}\) &&\(2\)&\(3\)&\(1\)&\(2\)&\(3\)&\(2\)&\(0\)&\(0\)&\(2\)&\(2\)&\(1\)   \\

 \(a_{3}\) &&&\(3\)&\(3\)&\(3\)&\(2\)&\(3\)&\(3\)&\(3\)&\(3\)&\(3\)&\(2\)   \\

 \(a_{4}\) &&&&\(1\)&\(3\)&\(3\)&\(2\)&\(3\)&\(0\)&\(3\)&\(3\)&\(2\)   \\

 \(a_{5}\) &&&&&\(3\)&\(1\)&\(1\)&\(2\)&\(2\)&\(2\)&\(3\)&\(1\)   \\

 \(a_{6}\) &&&&&&\(1\)&\(1\)&\(2\)&\(0\)&\(3\)&\(3\)&\(3\)   \\

 \(a_{7}\) &&&&&&&\(3\)&\(3\)&\(3\)&\(2\)&\(2\)&\(3\)   \\

 \(a_{8}\) &&&&&&&&\(3\)&\(3\)&\(2\)&\(3\)&\(3\)   \\

 \(a_{9}\) &&&&&&&&&\(3\)&\(2\)&\(3\)&\(1\)   \\

 \(a_{10}\)&&&&&&&&&&\(2\)&\(3\)&\(2\)   \\

 \(a_{11}\)&&&&&&&&&&&\(3\)&\(3\)   \\

 \(a_{12}\)&&&&&&&&&&&&\(3\)   \\

\hline
\end{tabular}
\end{center}

 The clustering problem can be considered as the following:

~~

 Find the clustering solution
  \( \widetilde{X} = \{X_{1},...,X_{\iota},...,X_{\lambda} \}) \)
  (i.e., a set of student teams as
  clusters/groups without intersection)
 for implementation of special laboratory work(s)
 (e.g., in radio engineering)
  while taking into account some requirements:

 (a) constraints for the number of elements (students)  in each cluster
 (i.e., cluster cardinality):~

 \( \eta_{min} \leq |X_{\iota}| \);  

 (b) constraint for total inclination/skill in the cluster:~
    \( \xi(X_{\iota}) \succeq  \xi_{0}=(2,3,3,2) \);

 (c) the objective function 1 is to minimize the balance index by
 cluster cardinality~
 \(B^{c} (\widetilde{X})\);

 (d) the objective function 2 is to maximize
 (multicriteria case)
 the  worst cluster structure estimate;

 (e) the objective function 3 is to maximize
  the  worst inter-cluster cluster estimate
 of element compatibility.

~~

 The version of the optimization problem for the example above
 can be considered as follows:
  \[\min~ B^{c} (\widetilde{X}) =
   \max_{\iota = \overline{1,\lambda}} ~|X_{\iota}|
 - \min_{\iota = \overline{1,\lambda}} ~|X_{\iota}|,
    ~~~~~
  \max \min_{\iota = \overline{1,\lambda}}  \xi (X_{\iota}),
  ~~~~~
 \max \min_{\iota = \overline{1,\lambda}} \varepsilon \ (X_{\iota})
 \]
 \[s.t. ~~~~
  3 \leq |X_{\iota}|\leq 4 ~~ \forall \iota = \overline{1,\lambda},
   ~~~~~ \xi(X_{\iota}) \succeq \xi_{0}=(2,2,3,2) ~~\forall \iota = \overline{1,\lambda} . \]
 The combinatorial optimization models of this kind are very complicated
 (i.e., NP-hard).
 Thus, enumerative algorithms or heuristics (metaheuristics) are
 used.

 Now, it is reasonable to describe a simplified heuristic (for the example above):

~~

 {\it Stage 1.} Counting an approximate number of clusters
 (e.g., \(4\)).

 {\it Stage 2.} Selection of the most important criteria:~
   1st choice: criterion 3,  2nd choice: criterion 2.

 {\it Stage 3.} Selection of the best elements
 (about \(4\))
 from the viewpoints
 of the selected criteria (e.g., by Pareto-rule)
 as kernels
 (``domain(s) leaders'')
 of the future clusters/teams.
 In general,
 small cliques or quasi-cliques can be considered as the kernels.
 In the example, the elements are:
 \(a_{1}\), \(a_{3}\), \(a_{6}\), \(a_{9}\).

 {\it Stage 4.} Extension of each kernel above by other elements while
 taking into account element compatibility.
 As a result,
 the following clustering solution can be considered:
  \( \widetilde{X} =\{X_{1},X_{2},X_{3},X_{4} \}  \)
  (\(B^{c} (\widetilde{X}) = 1\)), where
  the clusters are:

 (i) \(X_{1} = \{ a_{1},a_{2},a_{4} \}\),
  \(\xi(X_{1}) =(2,2,3,3)\),
    \( \varepsilon ( X_{1}) = 8 \);

 (ii) \(X_{2} = \{ a_{3},a_{7},a_{8} \}\),
  \(\xi(X_{2}) =(2,3,3,2)\),
    \( \varepsilon ( X_{2}) = 8 \);

 (iii) \(X_{3} = \{ a_{6},a_{5},a_{11} \}\),
   \(\xi(X_{3}) =(3,3,3,3)\),
    \( \varepsilon ( X_{3}) = 8 \);

 (iv) \(X_{4} = \{ a_{9},a_{10},a_{12},a_{13} \}\),
  \(\xi(X_{4}) =(3,3,3,3)\),
    \( \varepsilon ( X_{4}) = 15 \).

 Note, balance index by total  inter-cluster edge weight is
 \(B^{v} (\widetilde{X}) = 7 \).

~~

 Fig. 17 depicts the obtained four clusters solution
  \( \widetilde{X} =\{X_{1},X_{2},X_{3},X_{4} \}  \)
 (edge weight/compatibility estimates are pointed out).

~~

\begin{center}
\begin{picture}(28,25)
\put(32,00){\makebox(0,0)[bl]{Fig. 17. Clustering solution}}

\put(0.5,06){\makebox(0,0)[bl]{\(a_{1}\)}}
\put(16.5,06){\makebox(0,0)[bl]{\(a_{4}\)}}
\put(05.4,17){\makebox(0,0)[bl]{\(a_{2}\)}}

\put(05,12){\makebox(0,0)[bl]{\(2\)}}
\put(13.5,12){\makebox(0,0)[bl]{\(3\)}}
\put(09,05){\makebox(0,0)[bl]{\(3\)}}

\put(05,08){\circle*{1.5}} \put(05,08){\circle{2.2}}
\put(15,08){\circle*{1.5}} \put(10,18){\circle*{1.5}}

\put(05,08){\line(1,0){10}} \put(05,08){\line(1,2){05}}
\put(15,08){\line(-1,2){05}}

\put(02.5,20){\makebox(0,0)[bl]{Cluster \(X_{1}\)}}

\end{picture}
%
\begin{picture}(28,25)

\put(0.5,06){\makebox(0,0)[bl]{\(a_{3}\)}}
\put(16.5,06){\makebox(0,0)[bl]{\(a_{8}\)}}
\put(05.4,17){\makebox(0,0)[bl]{\(a_{7}\)}}

\put(05,12){\makebox(0,0)[bl]{\(2\)}}
\put(13.5,12){\makebox(0,0)[bl]{\(3\)}}
\put(09,05){\makebox(0,0)[bl]{\(3\)}}

\put(05,08){\circle*{1.5}} \put(05,08){\circle{2.2}}
\put(15,08){\circle*{1.5}} \put(10,18){\circle*{1.5}}

\put(05,08){\line(1,0){10}} \put(05,08){\line(1,2){05}}
\put(15,08){\line(-1,2){05}}

\put(02.5,20){\makebox(0,0)[bl]{Cluster \(X_{2}\)}}

\end{picture}
%
\begin{picture}(28,25)

\put(0.5,06){\makebox(0,0)[bl]{\(a_{6}\)}}
\put(16.5,06){\makebox(0,0)[bl]{\(a_{11}\)}}
\put(05.4,17){\makebox(0,0)[bl]{\(a_{5}\)}}

\put(05,12){\makebox(0,0)[bl]{\(3\)}}
\put(13.5,12){\makebox(0,0)[bl]{\(2\)}}
\put(09,05){\makebox(0,0)[bl]{\(3\)}}

\put(05,08){\circle*{1.5}} \put(05,08){\circle{2.2}}
\put(15,08){\circle*{1.5}} \put(10,18){\circle*{1.5}}

\put(05,08){\line(1,0){10}} \put(05,08){\line(1,2){05}}
\put(15,08){\line(-1,2){05}}

\put(02.5,20){\makebox(0,0)[bl]{Cluster \(X_{3}\)}}

\end{picture}
%
\begin{picture}(25,25)

\put(01,05){\makebox(0,0)[bl]{\(a_{9}\)}}
\put(20.6,05){\makebox(0,0)[bl]{\(a_{13}\)}}
\put(00,18.5){\makebox(0,0)[bl]{\(a_{10}\)}}
\put(21,17.5){\makebox(0,0)[bl]{\(a_{12}\)}}

\put(02.5,12){\makebox(0,0)[bl]{\(3\)}}
\put(21,12){\makebox(0,0)[bl]{\(3\)}}
\put(12,15.2){\makebox(0,0)[bl]{\(3\)}}
\put(12,05){\makebox(0,0)[bl]{\(1\)}}
\put(16,11){\makebox(0,0)[bl]{\(2\)}}
\put(07,11){\makebox(0,0)[bl]{\(3\)}}

\put(05,08){\circle*{1.5}} \put(05,08){\circle{2.2}}
\put(20,08){\circle*{1.5}} \put(05,18){\circle*{1.5}}
\put(20,18){\circle*{1.5}}

\put(05,08){\line(1,0){15}} \put(05,18){\line(1,0){15}}
\put(05,08){\line(0,1){10}} \put(20,08){\line(0,1){10}}
\put(05,08){\line(3,2){15}} \put(20,08){\line(-3,2){15}}

\put(04.5,20){\makebox(0,0)[bl]{Cluster \(X_{4}\)}}

\end{picture}
\end{center}

\section{Conclusion}

 The paper describes
 the author preliminary outline
 of various combinatorial balancing problems
 including new balance indices for clustering solutions,
 multicriteria combinatorial models and examples.
 It may be reasonable to point out
 some prospective future research directions:

 (1) special investigation of balance indices
 for balanced structures (degree of balance, etc.);

 (2) study of various balancing problems in combinatorial
  optimization
 (e.g., balanced knapsack-like problems,
 balanced allocation problems, balanced bin packing problems);

 (3) taking into account uncertainty in the balancing models;

 (4) examination of balancing clustering problems in networking
 (e.g., design of multi-layer communication networks,  routing);

 (5) study of augmentation approaches to balance structures
 while taking into account the improvement of the structures
 balance;
 and

 (6) usage of the described approaches in CS/engineering education.

\section{Acknowledgments}

 The research was done in
 Institute for Information Transmission Problems of Russian
 Academy of Sciences
 and supported by the Russian Science Foundation
 (grant 14-50-00150, ``Digital technologies and their applications'').


\end{document}